\documentclass[lettersize,journal]{IEEEtran}
\usepackage{amsmath,amsfonts}
\usepackage{algorithmic}
\usepackage{algorithm}
\usepackage{hyperref}
\usepackage{array}
\usepackage[caption=false,font=footnotesize,labelfont=rm,textfont=rm]{subfig}
\usepackage{textcomp}
\usepackage{stfloats}
\usepackage{url}
\usepackage{verbatim}
\usepackage{graphicx}
\usepackage{booktabs}
\usepackage{makecell}
\usepackage{multirow}
\usepackage{enumitem}
\usepackage{pifont}
\usepackage{cite}
\usepackage{color}

\usepackage{longtable}

\def\BibTeX{{\rm B\kern-.05em{\sc i\kern-.025em b}\kern-.08em
    T\kern-.1667em\lower.7ex\hbox{E}\kern-.125emX}}
\usepackage{balance}

\begin{document}

\title{Industrial Metaverse: Enabling Technologies, \\Open Problems, and Future Trends}

\author{Shiying Zhang, Jun Li, \emph{Fellow, IEEE}, Long Shi, \emph{Senior Member, IEEE}, \\ Ming Ding, \emph{Senior Member, IEEE}, Dinh C. Nguyen, \emph{Member, IEEE},\\ Wen Chen, \emph{Senior Member, IEEE}, and Zhu Han \emph{Fellow, IEEE}
\thanks{
This work is supported in part by the Key Technologies R\&D Program of Jiangsu (Prospective and Key Technologies for Industry) under Grants BE2023022 and BE2023022-2, in part by National Natural Science Foundation of China (NSFC) under Grant 62371239 and 62471204, and in part by Major Natural Science Foundation of the Higher Education Institutions of Jiangsu Province under Grant 24KJA510003.
This work is supported by Shanghai Kewei 22JC1404000 and 24DP1500500.
This work is partially supported by NSF ECCS-2302469, CMMI-2222810, Toyota. Amazon and Japan Science and Technology Agency (JST) Adopting Sustainable Partnerships for Innovative Research Ecosystem (ASPIRE) JPMJAP2326.
\textit{(Corresponding author: Jun Li, Long Shi)}

Shiying Zhang and Long Shi are with the School of Electrical and Optical Engineering, Nanjing University of Science and Technology, Nanjing 210094, China (e-mail: shiying.zhang@njust.edu.cn and slong1007@gmail.com).

Jun Li is with the School of Information Science and Engineering, Southeast University, Nanjing, 210096, CHINA. (email: jun.li@seu.edu.cn).

Ming Ding is with Data61, CSIRO, Eveleigh, NSW, Australia (e-mail:
Ming.Ding@data61.csiro.au).

Dinh C. Nguyen is with the Department of Electrical and Computer Engineering, University of Alabama in Huntsville, USA (e-mail: Dinh.Nguyen@uah.edu).

Wen Chen is with Department of Electronic Engineering, Shanghai Jiao Tong University, Shanghai 200240, China (e-mail: wenchen@sjtu.edu.cn).

Zhu Han is with the Department of Electrical and Computer Engineering at the University of Houston, Houston, TX 77004 USA, and also with the Department of Computer Science and Engineering, Kyung Hee University, Seoul, South Korea, 446-701 (e-mail: hanzhu22@gmail.com).

}}


\maketitle

\begin{abstract}
As an emerging technology that enables seamless integration between the physical and virtual worlds, the Metaverse has great potential to be deployed in the industrial production field with the development of extended reality (XR) and next-generation communication networks. The Industrial Metaverse is used for product design, production operations, quality inspection, and testing. However, there is limited understanding of the enabling technologies associated with it, including the specific industrial scenarios targeted by each technology and the potential migration of technologies from other domains to the industrial sector. This paper provides a comprehensive survey of the latest literature on the Industrial Metaverse. We first analyze its advantages for industrial production, then review key enabling technologies such as blockchain (BC), privacy-preserving computing (PPC), digital twin (DT), fifth/sixth generation mobile communication technology (5G/6G), XR, and artificial intelligence (AI), and explore how these technologies support different aspects of industrial production. We also present major challenges in the Industrial Metaverse, including privacy and security concerns, resource limitations, and interoperability constraints, along with existing solutions. Finally, we outline several open issues and future research directions.

\end{abstract}


\section{Introduction}\label{1}

The Metaverse is a nascent concept that presents a model of interconnection between the virtual and real worlds\cite{Zhao2022MetaversePF}. This digital system integrates both worlds into a cohesive system, providing users with a high degree of freedom to create content and modify modules as they wish, while also allowing different participants to define their own expressions. 
Regarding the conceptual description of the Metaverse, some scholars argue that ``the Metaverse is a collection of virtual spaces, constructed by a series of technologies including virtual reality (VR) and augmented reality (AR). People continuously create new content through smart wearable devices as input terminals." Others describe it as ``the latest stage of visual immersion technology, essentially serving as an online digital space, applied to various fields of production and life." Although the term ``Metaverse" does not yet have a precise definition in the industry, its fundamental technological framework has gradually matured and has been successfully applied in various fields, including education, healthcare, and industrial manufacturing. The Industrial Metaverse refers to the creation of shared virtual spaces enabled by Metaverse technologies to support multi-user, multi-device industrial scenarios for 3D modeling and immersive interaction.

Particularly, during the COVID-19 pandemic, many employees were forced to work from home. At the same time, continuous advancements in digital technologies have driven transformation. Many traditionally offline production activities have shifted online, allowing employees to operate remotely, similar to the concept of Digital Twin (DT) technology. DT creates virtual replicas of physical objects, systems, or processes that reflect the real-world status in real-time for monitoring, optimization, and prediction. These virtual models are tightly coupled with the real world, capable of simulating and providing feedback on the physical world. Unlike the simple mapping of DTs, the Industrial Metaverse is not merely a one-to-one replica of the real world into the virtual world. It also enables interactions and experiences that have not yet been realized in the real world. In the Industrial Metaverse, all virtual worlds and derived elements, along with digital entities in virtual spaces, can achieve online sharing, enhanced perception, and content generation. These virtual worlds, through highly immersive interactions, support the visualization, simulation, and optimization of industrial operations, allowing experiments and operations that cannot be replicated or surpassed in the real world to be conducted in a virtual environment. This provides richer and more efficient industrial application scenarios. For example, product performance can be tested through fault simulations at almost zero cost, creating experimental conditions that are impossible to replicate in reality, and extending operational capabilities through the linkage between the virtual and physical worlds. Compared to typical virtual worlds, the focus of the industrial virtual world lies in simulating industrial processes, manufacturing environments, or operational scenarios in the real world. The Industrial Metaverse places high demands on cost control in industrial manufacturing, requiring its design and implementation to consider cost efficiency and optimization, aiming to minimize deployment and maintenance costs. Additionally, the Industrial Metaverse involves collaboration between different enterprises in areas such as industrial production and supply chain management, requiring a high level of trust in the execution of interactions.
Table \ref{table_industrial} lists the relationships between DT, Metaverse, and Industrial Metaverse.
The key benefits of Metaverse are highlighted as follows: 
\begin{itemize}
\item \textbf{Low-cost Simulation:} 
The Metaverse can be used to simulate production processes in a cost-effective way. It can be used to implement various services by defining virtualized roles and scenarios, to create an orderly hierarchy in an enterprise virtual environment, and to generate content and simulation results that can be used directly to predict and optimize the situation in a real factory.
\item \textbf{Cross-regional Collaboration:} 
The Metaverse transcends geographical constraints and acquires data from diverse sensors and production lines, empowering manufacturers or disparate departments within the same manufacturing entity, geographically scattered, to collaborate on production, thereby enhancing production efficiency.
\item \textbf{Real-Time Participation through Avatars:} 
Human participants in industrial manufacturing processes are represented by corresponding virtual avatars in the Industrial Metaverse. These avatars ensure system stability and further optimize production processes through immersive observation and sensory feedback\cite{10477604}.
\item \textbf{Secure Interaction Assurance:} 
Through the integration of security technologies such as Non-Fungible Tokens (NFTs), the Industrial Metaverse can achieve secure protection of digital ownership, thereby aiding enterprises in preventing unauthorized user intrusions and data leaks during collaborative production processes.

\begin{figure*}[!t]
\center{\includegraphics[width=0.92\textwidth]{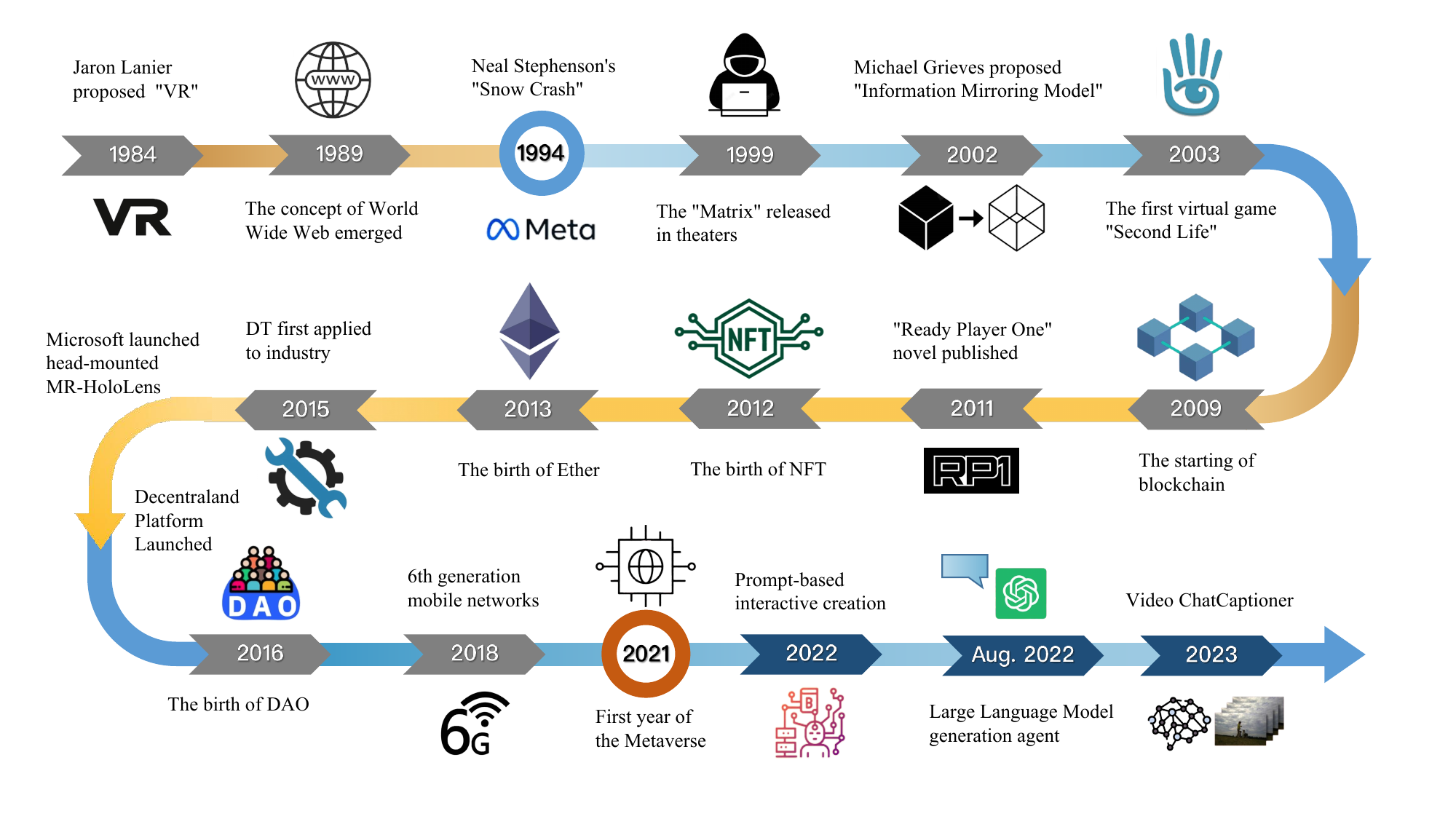}}
\caption{The development in the Metaverse from the perspective of technological updates. The concept of the Metaverse can be traced back to the VR concept proposed by Jaron Lanier in 1984. The formal concept of the Metaverse, however, was introduced in 1994. Over the following 30 years, enabling technologies and industries associated with the Metaverse (particularly the gaming industry) have continued to evolve. It was not until 2021 that there was a rapid surge in research achievements related to the Metaverse, which is why 2021 is also referred to as the ``Metaverse Year".
}
\label{fig1}
\end{figure*}

\end{itemize}

\begin{table*}[!t]
\begin{center}
\caption{Comparison of DT, Metaverse, and Industrial Metaverse.}
\label{table_industrial}
\renewcommand\arraystretch{1.3}
\begin{tabular}{|m{1.3cm}<{\centering}|p{3cm}|p{5.3cm}|p{3cm}|p{2.5cm}|}
\hline
\textbf{Aspect} & \textbf{Scope} & \textbf{Key Features} & \textbf{Real-time Requirement} & \textbf{User Interactivity}\\
\hline
DT & 
Simulation and predictive maintenance. & 
Optimize the performance of physical assets or systems through real-time data synchronization and simulation. &
High. &
Lower.\\
\hline
Metaverse  & 
Entertainment, social and digital economy, etc. & 
Focus on immersive experience and social interaction. &
Medium. &
Very high.\\
\hline
Industrial Metaverse & 
Industrial application. & 
In industrial operations, it supports data-driven decision-making and system optimization. &
Very high. &
Very high.\\
\hline
\end{tabular}
\end{center}
\end{table*}

In summary, the Industrial Metaverse, as a novel intelligent manufacturing paradigm, is poised to catalyze an unprecedented industrial upgrading, and its inherent value is expected to significantly exceed its consumption value. Fig. \ref{fig1} shows the development of the main enabling technologies in the Metaverse. 

\begin{table}[!t]
\begin{center}
\caption{Key Abbreviations.}
\label{table1}
\renewcommand\arraystretch{1.1}
\begin{tabular}{ c | c }
\hline
 \textbf{Abbreviation} &  \textbf{Explanation} \\
\hline
AI & Artificial Intelligence\\
\hline
AIGC & AI Generated Content\\
\hline
AOI & Age of Information\\
\hline
AR & Augmented Reality\\
\hline
ASP & AI Service Provider\\
\hline
BC & Blockchain\\
\hline
BIM & Building Information Modeling\\
\hline
CeAI & Centralized AI\\
\hline
CMS & Cyber-Manufacturing Systems\\
\hline
DeAI & Decentralized AI\\
\hline
DP & differential privacy\\
\hline
DT & Digital Twin\\
\hline
FL & Federated Learning\\
\hline
GANs & Generative Adversarial Networks\\
\hline
HCI & Human-computer Interaction\\
\hline
HE & homomorphic encryption\\
\hline
HMDs & Head-mounted Displays\\
\hline
IIoT & Industrial Internet of Things\\ 
\hline
IoT & Internet of Things\\
\hline
LPWA & Low-Power Wide-Area\\
\hline
LSTM & Long Short-Term Memory\\
\hline
MEC & Mobile Edge Computing\\
\hline
MR & Mixed Reality\\
\hline
MSF & Metaverse Standards Forum\\
\hline
NFT & Non-Fungible Token\\
\hline
O\&M & Operations \& Maintenance\\
\hline
PLC & Product Lifecycle\\
\hline
PPC & Privacy-Preserving Computing\\
\hline
PT & Physical Twin\\
\hline
QoS & Quality of Service\\
\hline
SemCom & Semantic Communication\\
\hline
TEE & Trusted Execution Environment\\
\hline
URLLC & Ultra-reliable and Low-latency Communication\\
\hline
VAEs & Variational Autoencoders\\
\hline
VR & Virtual Reality\\
\hline
VSPs & Virtual Service Providers\\
\hline
XR & Extended Reality\\
\hline
XRI & XR-IoT\\
\hline
\end{tabular}
\end{center}
\end{table}

\subsection{Related Works}\label{1B}
Neal Stephenson's novel ``Snow Crash" was the first to introduce the concept of the Metaverse, a digital landscape constructed through AR technology. Since then, there has been an explosion in literature on the Metaverse, with much of it focused on its fundamental characteristics\cite{Dionisio20133DVW}, empowering technologies\cite{Park2022AMT}, and privacy and security aspects. Some works also analyze the platform deployment and necessary components required for its application scenarios.
Our research on the industrial applications of the Metaverse is comprehensive and detailed in Table \ref{table2}. 

In terms of the empowering technologies within the Industrial Metaverse, the researchers in \cite{Mourtzis2023BlockchainII} delineate the technical architecture of blockchain (BC) in Metaverse Industrial 5.0, including its roles and limitations in data collection, storage, sharing, interoperability, and privacy protection. 
The work of \cite{Yang2022ApplicationOD} investigates the application of fluid mechanical components in numerical simulation and fault detection, supported by the Metaverse and DT.
Additionally, the article by \cite{Li2022WhenIO} expounds on the motivations and utilities of the four fundamental technologies of the Metaverse: AI, data communication, DT, and mobile edge computing (MEC). Furthermore, it describes seven crucial requirements for the current amalgamation of the Metaverse with the Internet of Things (IoT). 
Moreover, in \cite{Bhattacharya2023TowardsFI}, the authors present a cross-industry vertical Metaverse reference architecture, elucidating the underlying business logic of the contract interface as the backend, along with the front-end and back-end interaction mechanism implemented using the Web3.0 browser engine.

Although some of the works do not specifically focus on industrial applications of the Metaverse, the enabling technologies involved can be directly transferred to the manufacturing industry. 
For instance, Dionisio \emph{et al.} \cite{Dionisio20133DVW} provide a comprehensive summary of the four key characteristics of the Metaverse, namely, realism, ubiquity, interoperability, and scalability, which can be leveraged to improve manufacturing operations. 
Furthermore, Kye \emph{et al.}\cite{Kye2021EducationalAO} conduct a thorough analysis of the potential, as well as the constraints, of utilizing the Metaverse in educational contexts. This encompasses the domains of curriculum design and platform development, which could be effectively applied to skill training scenarios within the industry.

\begin{figure*}[!t]
\center{\includegraphics[width=0.97\textwidth]{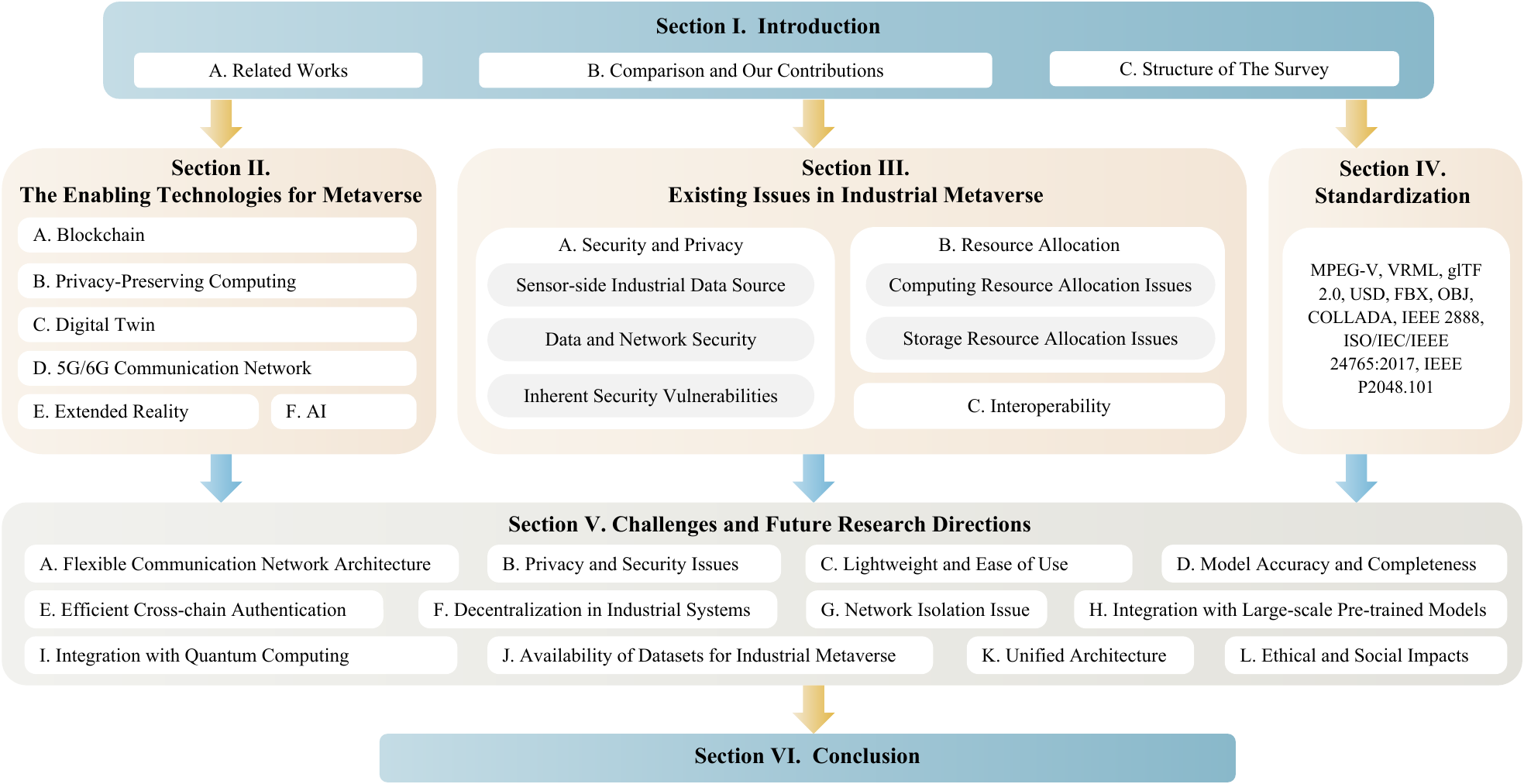}}
\caption{Structure of the survey.}
\label{fig2}
\end{figure*}

\subsection{Comparison and Our Contributions}\label{1C}

We have noticed that the Metaverse shows great potential in the industrial sector and has gradually attracted people's attention. Although there are numerous reviews on enabling technologies such as DT in industrial scenarios, research on Industrial Metaverse is scarce, mostly listing broad application scenarios without exploring the interconnection between enabling technologies or their roles in the architecture of Industrial Metaverse. 
In the work of\cite{Bhattacharya2023TowardsFI}, although the authors provide Metaverse architecture design for vertical industries, they do not discuss the production processes covered by enabling technologies in industrial production and analyze their specific technical support roles. The work of\cite{Tlili2023MetaverseFC}, although addressing certain industry application challenges, lacked comprehensiveness by omitting a discourse on the enabling technologies involved. Overall, at present, other research works have not formed a framework that matches the industrial production scenario, and there is a lack of in-depth combing and analysis from the perspective of industrial scenarios and technology applications.
Furthermore, existing research lacks a framework that matches industrial production scenarios and lacks in-depth analysis from the perspective of industrial scenarios and technological applications.
Additionally, compared to other application scenarios of the Metaverse, the industrial setting necessitates a high degree of real-time responsiveness to ensure synchronization and timely responses with the actual production environment. Any delay in the production environment could result in production line downtime or decreased efficiency. Additionally, interoperability in cross-enterprise collaboration is an indispensable feature of the Industrial Metaverse. Unlike typical commercial Metaverse scenarios, the equipment, systems, and software involved in industrial production usually originate from different vendors, utilizing diverse interface standards, communication protocols, and data formats. These elements are intertwined throughout a series of processes including design, production, testing, deployment, and feedback, each involving collaboration among various departments. Lastly, security and privacy concerns are paramount. Industrial production involves a plethora of sensitive data and confidential information such as design blueprints and industrial plans. Therefore, the Industrial Metaverse imposes stricter security requirements and must possess robust security mechanisms to ensure sensitive information is not leaked or maliciously tampered with. Hence, we believe it is necessary to comprehensively analyze and explore the application and deployment of Metaverse technology in industrial settings, and further advance related research endeavors.
To this end, this paper provides a set of key contributions as follows: 
\begin{itemize}
\item{Re-evaluate the architecture of the Industrial Metaverse, outlining its current research status and key enabling technologies, while delineating the interactions among these enabling technologies and their roles within the architecture.}
  
\item{Summarize existing research achievements on Metaverse enabling technologies and their application status in various stages of industrial production, identifying existing deficiencies for each enabling technology.}

\item{Analyze pressing issues in the current stage of the Industrial Metaverse, including privacy concerns, resource allocation, and platform interoperability, while examining existing solutions and shortcomings.}

\item{To advance industrial deployment, we summarize the current standardization efforts and open challenges in the Industrial Metaverse, and provide insights into the future development and research directions of the Industrial Metaverse, offering guidance for future work.}

\end{itemize}

To the best of our knowledge, this is the first comprehensive work to organize and analyze the Industrial Metaverse and its core enabling technologies.

\begin{table*}[!t]
\begin{center}
\caption{Survey papers of Industrial Metaverse.}
\label{table2}
\renewcommand\arraystretch{1.2}
\begin{tabular}{|m{3cm}<{\centering}|c|p{3.8cm}|p{2.2cm}|p{6cm}|}
\hline
\textbf{Ref.} & \textbf{Year} & \textbf{Scope} & \textbf{Technologies} & \textbf{Key Contents}\\
\hline
Li \emph{et al.}\cite{Li2022WhenIO} & 2022 & Socialization, healthcare, education, smart city, entertainment, real estate & AI, 5G/6G, DT, MEC & The authors describe the motivation and utility of using the four Metaverse pillar technologies.\\
\hline
Yang \emph{et al.}\cite{Yang2022ApplicationOD} & 2022 & Fluid machinery pumps and fans & DT, BC, AI, XR & The authors investigate the application of fluid mechanical components supported by Metaverse and DT in the field of numerical simulation and fault detection.\\
\hline
Chang \emph{et al.}\cite{DBLP:journals/jcin/ChangZLXGSXKNQW22} & 2022 & Education, product testing, telecommuting, production optimization, smart cities & Computer vision, DT, BC & Three new Metaverse architectures at the edge, and technologies that help the Metaverse interact and share digital data.\\
\hline
Bhattacharya \emph{et al.}\cite{Bhattacharya2023TowardsFI} & 2023 & Manufacturing, internet-of-senses, industry, education, vehicle-to-everything, internet-of-bio-things & Web 3.0, 6G, BC, NFTs, XR & The authors discuss generalized Metaverse frameworks for modern industrial cyberspace.\\
\hline
Mourtzis \emph{et al.}\cite{Mourtzis2023BlockchainII} & 2023 & Data acquisition, storage, sharing, interoperability & BC, DT & The authors describe the technical architecture components of BC in Metaverse Industry 5.0.\\
\hline
Said \emph{et al.}\cite{Said2023MetaverseBasedLO} & 2023 & None & XR & This work illustrates the complexity of Metaverse online learning from the perspective of HCI.\\
\hline
Kshetri \emph{et al.}\cite{DBLP:journals/itpro/Kshetri23} & 2023 & Economics & XR, AI, 6G & The global economic impact of the Industrial Metaverse.\\
\hline
Dong \emph{et al.}\cite{DBLP:journals/iotm/DongZCJL23} & 2023 & Task-oriented communications & SemCom & Specific requirements for the three main types of tasks in the Industrial Metaverse.\\
\hline
\emph{Our paper} & 2023 & Industry 5.0 (Industrial data collection and storage, product design, operation training, system manufacturing, industrial quality control, etc.) & BC, DT, 5G/6G, XR, AI & An extensive survey of the Industrial Metaverse. Particularly,
\begin{itemize}
\item For the first time, we extensively discussed the role of several key enabling technologies at different stages of industry.
\item We comprehensively analyze the main existing problems and solutions in the Industrial Metaverse.
\item We first emphasize more comprehensive open challenges and research directions for the characteristics of industrial scenarios.
\end{itemize} \\
\hline
\end{tabular}
\end{center}
\end{table*}

\subsection{Structure of The Survey}\label{1D}
The sections of this paper are structured as follows. Section \ref{1} presents an overview of the fundamental principles underlying the Metaverse and elucidates the rationale behind its integration into industrial contexts. Additionally, it provides an inventory of pertinent Metaverse literature reviews, comparing them to identify the innovative aspects of our current undertaking. Section \ref{2} discusses around the five core enabling technologies of the Metaverse and summarizes the industrial applications and existing problems of each technology. Section \ref{3} succinctly summarizes the four primary challenges encountered in the Industrial Metaverse. In addition, in Section \ref{4} we collate some of the existing standardization efforts and platforms for the Industrial Metaverse. Moving forward, Section \ref{5} offers a comprehensive compilation of various unresolved hurdles prevailing in the contemporary Industrial Metaverse. Finally, our discourse culminates with a conclusive remark in Section \ref{6}.
Table \ref{table1} indicates the list of abbreviations employed in this paper.
Fig. \ref{fig2} outlines the organization of the survey.

\begin{figure*}[!t]
\center{\includegraphics[width=0.9\textwidth]{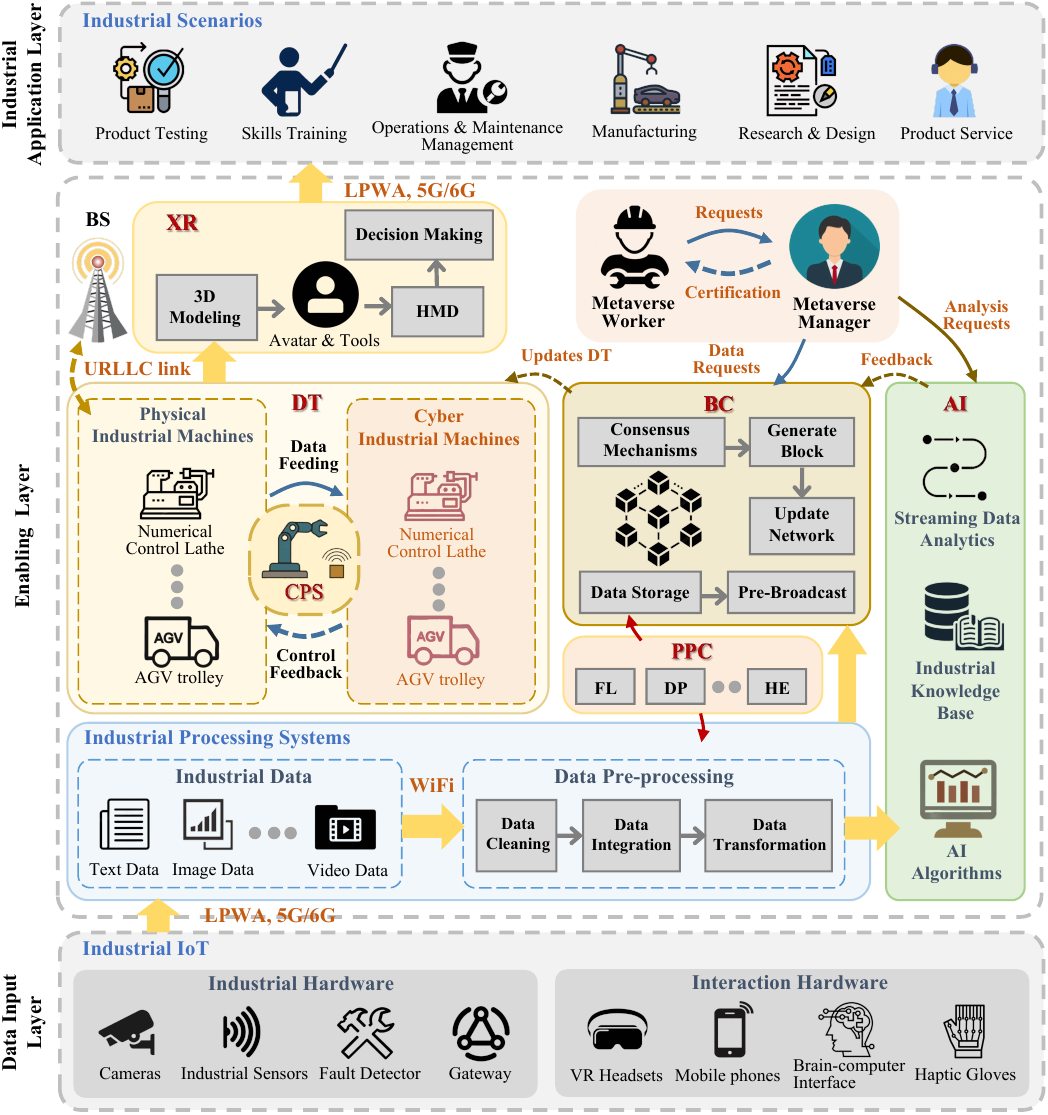}}
\caption{Industrial Metaverse Architecture. It consists of three layers: the data input layer, the enabling layer, and the industrial application layer. The enabling layer includes six components: AI, DT, BC, XR, the Metaverse management center, and the data processing system. In this context, the data input layer is composed of various industrial IoT devices. The collected multimodal industrial data is transmitted to the data processing system for preprocessing. The processed data can be stored in the BC or streamed to the AI component for analysis. The results of data analysis are used to respond to requests from the Metaverse management center, and decision outcomes are fed back to the BC. The BC performs updates on the network and DT. The DT achieves proportional replication and synchronization of virtual and physical devices and transfers parameter information to XR for 3D modeling. XR displays decision solutions through human-machine interaction devices and is applied in various industrial scenarios.
Among them, PPC ensures that data is not disclosed during processing and analysis.}
\label{fig3}
\end{figure*}

\section{The enabling technologies for Metaverse}\label{2}
This section introduces several core enabling technologies for the Industrial Metaverse, including BC, PPC, DT, XR, AI, and 6G. The primary roles of these enabling technologies in the Industrial Metaverse are as follows:

\begin{itemize}
\item{\textbf{DT:}  DT provides virtual representations of physical entities and precise descriptions of products. These fundamental data are provided by industrial sensors and cameras deployed within the enterprise. Combined with AI and autonomous robotics, DT in the Industrial Metaverse is used to represent physical systems in virtual space and meet various constraints and specifications. Simulation, as a critical component of the Metaverse, is used to test vast scenarios to provide optimal modeling and decision-making.}

\item{\textbf{AI:} Integrating AI technology can enhance the modeling accuracy of DT. AI can rapidly classify and analyze large amounts of data generated from factories, supply chains, or other production processes, aiding in pattern recognition and decision-making to adjust production processes, improve production quality, and achieve the intelligence and automation of production processes.}

\item{\textbf{XR:} VR and AR serve as immersive tools providing visualization, which are more intuitive, user-friendly, and easier to interpret compared to numbers in tables or points on charts. For example, Rockwell's development team built test scenarios on the Vuforia Studio platform, incorporating CAD files required for training tests to create wiring diagrams mapped to different product layouts and wiring schematics. Leveraging AR technology, Rockwell Automation achieved a 5\% reduction in training time and extended the same training method to other production lines.}
\end{itemize}

If DT, XR, and AI are prerequisites for enterprise decision-making and intelligent production in the Industrial Metaverse, then 5G/6G, BC, and privacy-preserving computing (PPC) provide the necessary low-latency and trustworthy interaction architecture.

\begin{itemize}
\item{\textbf{5G/6G:} Metaverse companies use 5G/6G networks to send and receive massive data on their Industrial Metaverse platforms. For instance, when an alert is received at the mission control center, operators can use 5G/6G connections to remotely view and operate robots via cameras. Furthermore, researchers at Nokia Bell Labs believe that compared to 5G, 6G will halve average power consumption and support a 10-fold increase in peak capacity.}

\item{\textbf{BC:} As a trusted distributed architecture, BC effectively enables trusted interaction among multiple enterprise platforms and facilitates identity authentication and rights assertion when several stakeholders (such as manufacturers, suppliers, and logistics providers) need to share data without a single party having control. Compared to traditional (distributed) databases combined with centralized management, BC may offer significant advantages, and in many Industrial Metaverse scenarios, it is also indispensable.}

\item{\textbf{PPC:} 
Lastly, at the data level, PPC technologies such as federated learning, differential privacy (DP), and homomorphic encryption (HE) can ensure that sensitive data is processed and analyzed without being disclosed. This enables multiple parties to share industrial data and collaborate on modeling and optimization while safeguarding privacy.}
\end{itemize}

In addition, we have investigated the latest advances in theoretical research and the development of Industrial Metaverse platforms, as well as the application of each enabling technology in different phases of industrial production. In a short summary section for each technology, we briefly summarize some of the issues that still need to be addressed when integrating these technologies with the Metaverse, as well as the technical requirements in industrial scenarios. 
Table \ref{table4} lists recent research works focused on enabling technologies for the Industrial Metaverse.

Additionally, we have depicted the reference architecture of the Industrial Metaverse in industrial scenarios, as shown in Fig. \ref{fig3}. It includes three layers: the data input layer, the enabling layer, and the industrial application layer. The enabling layer consists of six components: AI, DT, BC, XR, the Metaverse management center, and the data processing system.

Firstly, the data input layer comprises various IoT devices used to collect data sources in the factory, such as cameras, sensors (e.g., acoustic/light/humidity sensors), and user-facing human-machine interaction devices like VR headsets and mobile phones. The collected industrial data includes text data, image data, video data, and other modalities. These multimodal data sources are transmitted to the data processing system for preprocessing, including data cleansing, integration, and transformation. They are stored in the BC through a BC gateway for pre-broadcasting. On the other hand, workers in the Industrial Metaverse can initiate data upload requests themselves. The BC packs and encapsulates the validated data through smart contracts, generating new blocks for continuous updates. When worker submits a data analysis request, they first need to send the request to the administrator. After preliminary authentication by the administrator, the request message is forwarded to the AI component for data analysis\cite{Li2021BlockchainAD}. Meanwhile, the BC is responsible for recording transactions. The feedback from the data analysis is then used to update the DT. The DT subsequently sends it to the XR layer for 3D modeling, including modeling of production devices and environmental elements. The decisions obtained are displayed through XR headsets and serve various industrial scenarios in the industrial application layer, including product testing, skills training, and manufacturing. It is worth noting that the data used to update and adjust the DT model can come from the BC or directly from industrial equipment sensors, internal databases, or manually entered by operators\cite{Deng2024TrustworthyDP}. Through the DT component, the real world can be accurately replicated in digital form on multiple levels of virtual planes and can dynamically interact and synchronize with virtual devices on the virtual plane.

PPC ensures that during data transmission and analysis, privacy is maintained, thus allowing for secure data sharing and collaborative modeling across various participants in the Industrial Metaverse.
The communication between these layers and components is based on 6G, Low-Power Wide-Area (LPWA) and WiFi communication network to provide the required QoS for the exchanged data. LPWA is a low-power wide area network technology specialized in long range communication for IoT devices. In the Industrial Metaverse, LPWA can support the connection of sensors and IoT devices to the virtual environment. It can be used to monitor and collect data in the real world and transmit this data to the Industrial Metaverse for users to analyze and utilize.
  
In the ensuing discourse, we shall expound upon each of these constituent strata comprising the foundational edifice.

\subsection{Blockchain}\label{2A}

In the realm of the Industrial Metaverse, nodes of the Industrial Internet of Things (IIoT) amass vast quantities of sensor data in order to achieve instantaneous correspondence between the virtual and physical domains. Nevertheless, this pursuit carries with it the hazard of security and privacy breaches, as the information of these nodes is profoundly susceptible to compromise\cite{Li2021BlockchainAD}. Moreover, there exists a necessity to accomplish cross-platform interoperability within the Metaverse, grounded in a specific framework.

\begin{figure}[!t]
\center{\includegraphics[width=0.5\textwidth]{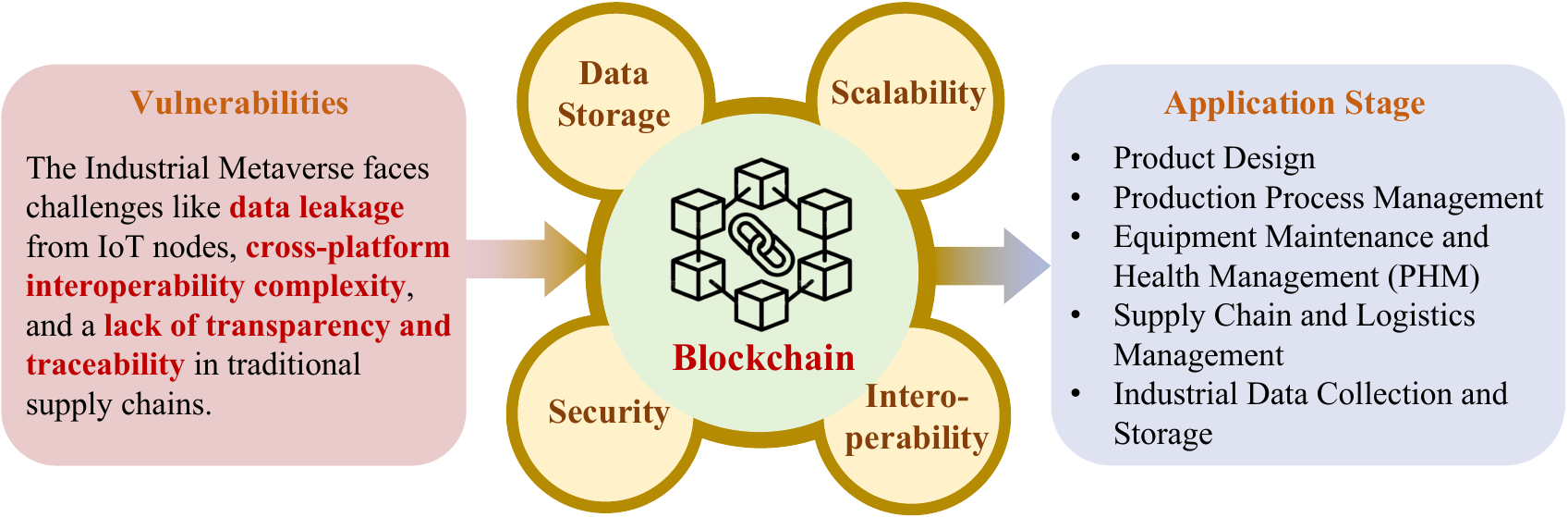}}
\caption{The vulnerabilities, advantages, and application stage of BC in Industrial Metaverse.}
\label{figBC}
\end{figure}

BC, as the fundamental technology of the Metaverse, is well-suited to meet the aforementioned requirements. Firstly, it leverages identity verification and consensus mechanisms to ensure the privacy and security of users, as well as the integrity of vast amounts of industrial data, while also providing a comprehensive transaction audit trail. Secondly, due to the decentralized nature of BC, it enables collaborative production among multiple manufacturers, allowing managers to schedule and coordinate activities across multiple platforms without the need for third-party verification platforms\cite{DBLP:journals/network/DuLSWWH23}.
Fig. \ref{figBC} shows the vulnerabilities, advantages, and application stage of BC in
Industrial Metaverse.

\subsubsection{Research Advances of Blockchain Enabling Industrial Metaverse}
Initially, the question of safeguarding privacy in the Metaverse was explored in the financial sector, where individuals use NFTs as digital property markers to attain secure data correspondence. Each NFT is linked to an owner, and ownership records are retained on a server. Nonetheless, due to the more intricate real-time information interaction and the heavy workload associated with production tokens in the context of industrial manufacturing scenarios, it has become necessary to establish a collaborative governance framework among enterprises that is composed of more comprehensive components and architecture.
To tackle this predicament, Lin \emph{et al.} \cite{Lin2022TowardsMM} propose a value-oriented, trusted, collaborative governance architecture as a blueprint for the deployment of the Metaverse in intelligent manufacturing scenarios. The authors provide a thorough summary of the functions and benefits of BC at various system levels, such as trustworthy data storage, secure transactions, and regulation.

It is noteworthy that computing resources, functioning as integrated asset exchanges, are also included in the domain of the Industrial Metaverse. However, the aforementioned architecture does not account for the issue of energy constraints caused by IIoT nodes. When devising an integration method between the Industrial Metaverse and BC, resource optimization and sharing should be fully taken into account. For instance, an enterprise management platform can employ various communication methods to offer computing power and encapsulate the communication steps of factory-deployed devices into a protocol. In this way, less powerful computing devices can utilize the support of stronger computing power.

The work presented in \cite{Xu2022MetaverseNC} introduces a communication paradigm for the native Metaverse, where users can access using encrypted addresses. A dedicated frequency band in the regional spectrum is utilized for the transmission of BC links in public domains. The controller can categorize access to specific resources to eliminate integration barriers between the Metaverse and inter-chain links. However, this encrypted address does not provide precise network topology information, and industrial IoT devices may possess mobility features, leading to a constantly changing topology structure that makes it impossible to determine the connection mode between nodes. Another practical concern is that profit-driven manufacturers typically consider whether the additional benefits of industrial synergy can cover participation costs. A potential solution is to design incentive mechanisms to attract companies to participate in routing-related activities. To address these two issues, the authors of \cite{Kang2022BlockchainbasedFL} propose a layered BC architecture with a main chain and multiple sub-chains for sensor data analysis, combined with federated learning (FL) and cross-chain technology. Additionally, an age-based smart contract is utilized as an incentive mechanism to achieve trustworthy model aggregation and updates.
Regarding the architecture design for dynamic topology scenarios, the authors of \cite{Wang2022BlockchainbasedER} adopt another approach for the architecture design, based on MEC. This solution allows MEC servers to possess additional resources for resource sharing and optimized utilization in non-trusted environments. This framework takes into account the heterogeneity of available server resources and user task requests, and utilizes a task assignment mechanism for edge resource sharing.
This BC-based MEC deployment scheme can significantly enhance the utilization of computing resources while also ensuring system security\cite{Du2023BlockchainAidedEC}.

Numerous studies have examined distinct scenarios for amalgamating the Metaverse with BC, such as architectural modeling and remote healthcare. Though these scenarios may not be relevant to our current discussion, they may potentially be customized to the industrial production domain, such as in the modeling of production devices and remote equipment control. For instance, Building Information Modeling (BIM) technology, which is utilized in 3D architectural modeling, can be exploited through parameterization methods to facilitate speedy data computation and storage for project management processes.
After exploring the components of BIM, the application of the Metaverse in virtual world construction, and the latest BC research, the authors of\cite{Huang2022FusionOB} put forward a preliminary BC-based BIM data trading framework. Moreover, the author also accentuates several potential future research topics, such as immersive design experiences in the Metaverse, the evolution, and retrospection of architectural scenarios.
Bhattacharya \emph{et al.}\cite{Bhattacharya2022MetaverseAT} raise the issue of security attacks in remote surgery, which can result in the uncontrollability of the mobile robot arm.

In general, the exploration of synergizing the Metaverse with BC is still in its nascent phase, with applications primarily focused on digital asset management, smart contracts, and decentralized identity verification to enhance the security and sustainability of the Metaverse.

\subsubsection{Application Stage in Industrial Production}

The integration of BC in industrial applications can bring forth a multitude of benefits. One such application is in the domain of supply chain and logistics. Raw materials need to circulate through different stages of production to enhance their value, and this mobility is dependent on the industrial supply chain. As an enabling technology for the Industrial Metaverse, BC can be utilized to track the entire lifecycle from raw materials to finished products, addressing issues such as the lack of traceability and low transparency in traditional supply chains\cite{RajaSanthi2022InfluenceOB}. Operators can monitor product flows and status changes in real-time, enhancing supply chain transparency and controllability, reducing human intervention errors, and providing a secure and efficient management model for product circulation.

In industrial data collection, the non-repeatability of data is crucial for accurate data analysis and effective business decisions. With BC, the hash value of each data block can be calculated and compared with the previous block's hash value to ensure that the collected data is not duplicated, thereby improving the reliability of data collection.

In terms of data storage, the massive amount of multimodal data generated during industrial production processes can quickly exhaust physical storage capabilities\cite{Mourtzis2023BlockchainII}. Overreliance on centralized storage systems in the Metaverse can result in a single point of failure, leading to significant economic loss. BC's decentralized paradigm can effectively solve the problem of single point of failure while preventing data tampering through the continuous creation of new blocks\cite{Shi2021PoolingIN}.

Furthermore, BC can achieve interoperability within virtual worlds through cross-chain protocols, simplifying connections using cross-chain technology. At the device level, it enables cross-chain transactions and value flows between different devices\cite{Deng2021BlockchainAF}. At the enterprise level, it facilitates collaboration among suppliers, manufacturers, and end-users, as well as cross-chain governance among different companies during the production stages. This reduces the complexity of system and enterprise integration, enhances operational efficiency, and allows operators to focus more on core tasks.

\subsubsection{Summary}
BC can enable efficient management of Industrial Metaverse platforms by leveraging its characteristics of immutability, traceability, collective maintenance, and transparency, thus reducing operational risks and human errors\cite{Wan2022BlockchainAA}. However, in a BC network, each participating node needs to validate transactions, a process that involves network transmission, disk writing, and other operations, leading to potential delays\cite{Shi2024BlockchainAidedDT}. For example, during the validation process, if network latency affects the real-time synchronization of production data and task response speed, operators may face information lags, which could lead to erroneous decisions or production interruptions. While some studies have proposed optimizing network topology, designing new consensus algorithms, and adopting hierarchical structures to alleviate this issue, such solutions are still limited by network scale and the number of nodes involved. Additionally, in some industrial scenarios, trust relationships between participants have already been established or can be guaranteed through centralized authentication systems. For example, in a private industrial network led by large enterprises or multiple partners, all participants may already trust a centralized management platform or permission system.

\subsection{Privacy-Preserving Computing}\label{2B}
\begin{figure}[!t]
\center{\includegraphics[width=0.5\textwidth]{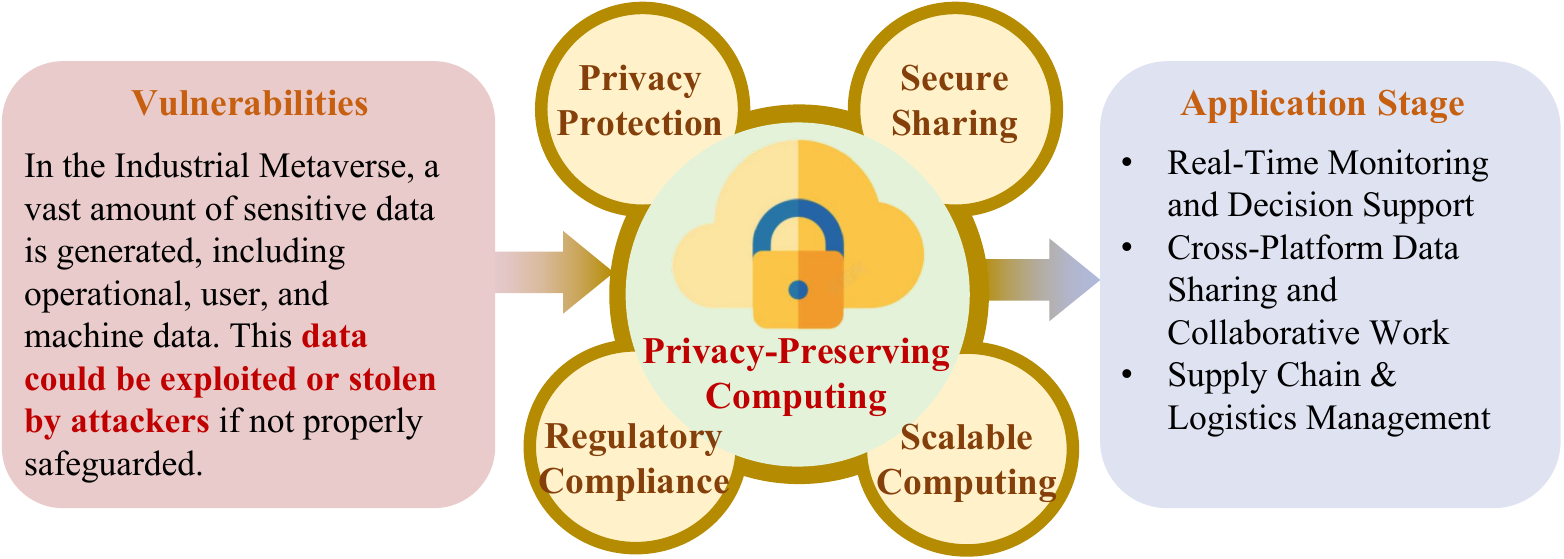}}
\caption{The vulnerabilities, advantages, and application stage of PPC in Industrial Metaverse.}
\label{figPPC}
\end{figure}

In the context of the Industrial Metaverse, blockchain primarily relies on its immutability and consensus mechanisms to ensure the secure operation of the platform, while PPC focuses on the protection of sensitive information during cross-platform and cross-domain data sharing and computation. Technologies such as FL and DP enable secure data sharing and computation without disclosing users' private and sensitive information, thus ensuring the security of data for both users and enterprises\cite{Wei2023PersonalizedFL}.
Fig. \ref{figPPC} shows the vulnerabilities, advantages, and application stage of DT in
Industrial Metaverse. 

\subsubsection{Research Advances of PPC Enabling Industrial Metaverse}
FL, as a classic distributed machine learning method in the field of PPC, effectively protects data privacy by allowing data to be stored on local devices, avoiding centralized storage and data exchange. Each participant performs local computation, model training, and updates, then aggregates model parameters to the central server instead of exchanging raw data. This enables collaborative training of machine learning models without sharing sensitive data. FL in the Metaverse has been regarded as a promising method because it ensures that raw data remains local to users while also reducing the memory and computational demands on the central server by integrating FL into the Metaverse\cite{Chen2023FederatedLF}.

However, despite FL’s ability to aggregate data without directly uploading raw data, malicious clients could still indirectly degrade the accuracy of the global model in the Metaverse by uploading erroneous model parameters. Sun \emph{et al.}\cite{Sun2024FedKCPF} point out that although some research has implemented defense mechanisms to prevent server attacks during the aggregation of global models, these defenses do not mitigate the effects of high-intensity attacks. Additionally, variations in device performance lead to differences in data feature distributions and quality. When there are significant differences in participants' data distributions, the global model has difficulty converging and suffers from poor performance. To address this, they propose a client-based defense mechanism that combines KNN with deep neural network feature representations to facilitate the rapid improvement of local model performance. Marios \emph{et al.}\cite{Aristodemou2024BayesianOP} further explore personalized FL (PFL) and proposed an adversarial poisoning attack model to evaluate PFL's susceptibility. By utilizing Bayesian optimization for layer optimization and optimizing the hidden layers of local models based on mutual information and Kullback-Leibler divergence, they aim to reduce classification uncertainty in the global model within PFL.

It is important to note that, in addition to poisoning attacks, attackers might also analyze aggregated model parameters to infer individual participants' training data, potentially recovering sensitive user information. Therefore, HE has become an indispensable privacy-preserving technique in FL, as it allows computations to be performed on encrypted data without decrypting it.

However, FL in the Metaverse, which handles large-scale data, requires efficient and lightweight security and privacy solutions. Security and efficiency must be considered simultaneously. While HE is a reliable security mechanism, it is not suitable for large-scale industrial data scenarios due to its computational cost. FL must focus on how to protect data and from what perspective. For example, the random response method in FL is efficient, but it does not always handle the complex data generated by Metaverse devices. Some research has proposed lightweight HE methods to address this issue. For instance, Du \emph{et al.}\cite{Du2022ALH} introduce a lightweight HE method based on erroneous formalism learning, where the original data is encrypted as a vector and a key exchange matrix computes the final ciphertext. However, current lightweight HE methods still face efficiency challenges when processing large-scale data and lack sufficient flexibility for diverse application scenarios.

In this context, DP serves as a complementary privacy-preserving solution. It ensures that attackers cannot infer any specific individual’s private information from the model's output by adding noise to the data during processing and analysis. Compared to HE, DP incurs lower computational overhead and is suitable for large-scale data processing and real-time systems. However, the introduction of DP in the Metaverse applications often leads to a trade-off between privacy and utility. The key research challenge in applying DP is minimizing the loss of data utility while ensuring privacy protection. Liu \emph{et al.}\cite{Liu2023WassersteinGA} construct a basic model for DP-based Metaverse data sharing using WGAN (Wasserstein GAN) and achieved a balance between privacy and utility by incorporating appropriate regularization and Wasserstein distance. Letafati \emph{et al.}\cite{Letafati2023GlobalDP} consider global DP in distributed Metaverse systems, drawing inspiration from the "invisibility mode" concept, and achieved a balance between diagnostic accuracy and loss function under different privacy levels.

As the complexity of the Metaverse environment continues to grow, large language models (LLMs), as powerful natural language processing tools, are beginning to show great potential in Industrial Metaverse. Lyu \emph{et al.}\cite{lyu-etal-2020-differentially} propose a method called Differential Private Neural Representation (DPNR), which further enhances privacy protection by discarding sensitive data. However, their focus was primarily on the inference phase, neglecting privacy issues during training. To address this, Qu \emph{et al.}\cite{Qu2021NaturalLU} apply local DP mechanisms to fine-tune the BERT model, balancing privacy protection with model utility. Furthermore, the MA framework proposed by Abadi \emph{et al.}\cite{Abadi2016DeepLW} and the privacy budget algorithm by Meiser and Mohammad provide theoretical methods to calculate the privacy cost of DP, further enhancing privacy protection effectiveness.

In addition to HE and DP, other privacy-preserving technologies such as Secure Multi-Party Computation (SMPC) and Zero-Knowledge Proofs (ZKP) are also continually evolving and provide alternative solutions for data privacy protection. SMPC allows multiple participants to jointly compute without revealing their private data, while ZKP enables one party to prove the truth of a statement without disclosing the underlying data. However, their practical applications in the Industrial Metaverse are still in the early stages, with relatively few research efforts.

\subsubsection{Application Stage in Industrial Production}
In the early stages of the Industrial Metaverse, vast amounts of data are generated by various devices, sensors, and users, which often contain sensitive information. PPC technologies, such as FL, enable multiple participants to collaboratively perform computations and training tasks without disclosing their individual data, effectively protecting data privacy. 

In the process of data sharing and collaboration, privacy-preserving technologies are widely applied to prevent the leakage of private information from individual users. For instance, when model parameters are shared among multiple devices or organizations, DP adds noise to ensure that attackers cannot infer specific user data from the model's output, thereby effectively protecting privacy. Meanwhile, zero-knowledge proofs allow one party to prove the authenticity of certain data or information without disclosing the actual content, providing additional security in the process of data sharing and validation.

\subsubsection{Summary}
Overall, with the continuous development of privacy-preserving technologies, data sharing and computation in the Industrial Metaverse will become more secure and efficient. By integrating FL, HE, DP, and other technologies, it will be possible to achieve cross-platform and cross-domain data sharing and collaboration without compromising privacy protection. Future research may focus on further optimizing these technologies, enhancing their efficiency in large-scale and real-time data processing, while maintaining effective privacy protection and reliability.

\subsection{Digital Twin}\label{2C}
DT can produce a virtual replica of a physical factory by gathering and analyzing production data and equipment parameters in the actual facility. With this information, DT can perform behavior predictions of physical objects or machine states. By accurately mirroring the real world in the virtual world, DT can resolve complex issues that would otherwise prove challenging to solve.

\begin{figure}[!t]
\center{\includegraphics[width=0.5\textwidth]{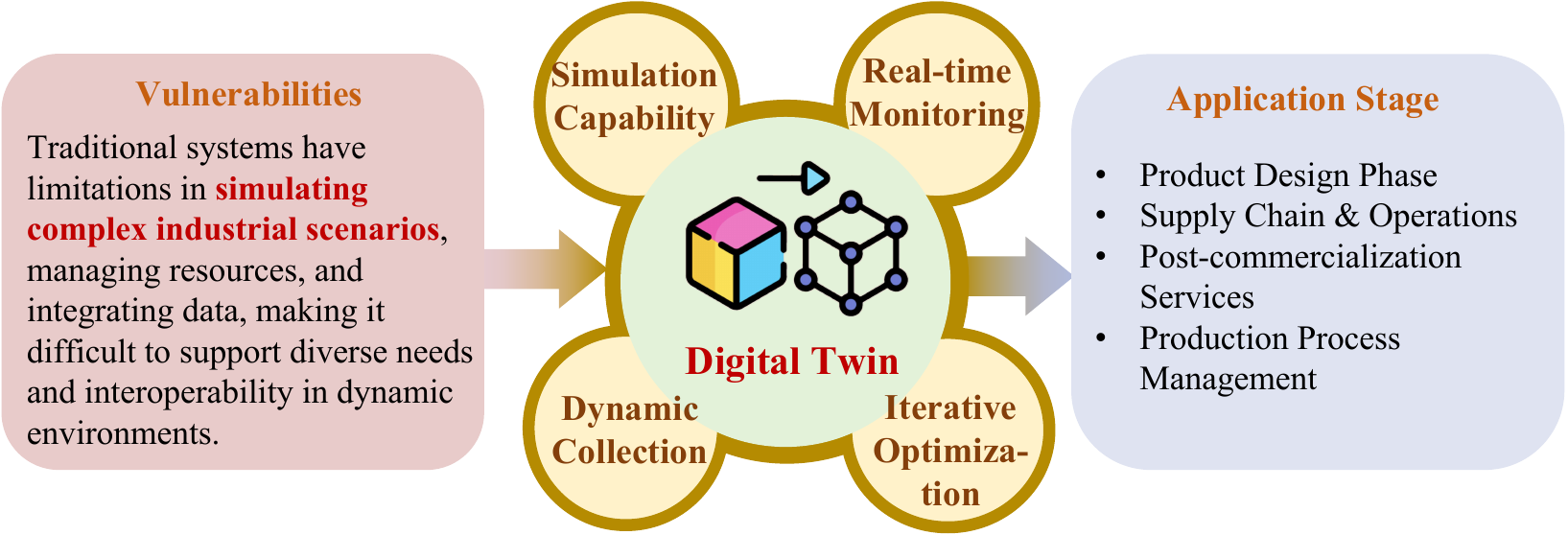}}
\caption{The vulnerabilities, advantages, and application stage of DT in Industrial Metaverse.}
\label{figDT}
\end{figure}

Hence, to develop an Industrial Metaverse platform, DT must fulfill three vital requisites: the capability to simulate industrial entities, the capacity to dynamically procure data from the live environment, and the proficiency to synchronize the data in real-time.
Fig. \ref{figDT} shows the vulnerabilities, advantages, and application stage of DT in
Industrial Metaverse. 

\subsubsection{Research Advances of DT Enabling Industrial Metaverse}
DT made its debut in the industrial sphere in 2003, when Michael Grieves proposed its implementation in product lifecycle (PLC) management. By utilizing production data and digital models, simulated processes became possible. As the IIoT and DT evolved, the primary focus shifted towards using DT to reduce industrial sites and enhance product evaluation through assistive technologies.

However, some researchers argue that DT tends to overlook crucial action details and treat operators and small devices as mass points in simulations, affecting object modeling. To overcome this challenge, Zhou \emph{et al.}\cite{Zhou2022IntelligentSO} design a small object detection framework that fuses multidimensional sensor data, treating devices, products, and operators as fundamental environmental parameters. They implemented multi-level feature fusion based on hybrid neural networks created with MobileNetv2, YOLOv4, and Openpose.
To effectively reduce the false detection rate, Wu \emph{et al.}\cite{Wu2022DigitalTO} utilize multi-modal fusion data as a data source, Cyber-Manufacturing Systems (CMS) interface for numerical insertion and visualization in DT for product surface defect detection tasks. Finally, they employed morphological alignment algorithms. Numerous studies have revealed that improving the simulation capability of DT requires integrating and optimizing AI algorithms.

Furthermore, given that DT necessitates substantial bandwidth and computing resources, it inherently demands real-time synchronization with stringent reliability and low latency prerequisites\cite{Guo2021SynchronizationOS}. As a result, achieving real-time synchronization for DT entails stringent reliability and low latency requirements. Therefore, an important research topic is the deep integration of DT into industrial platforms. This includes designing a shared architecture for digital factories and developing scheduling algorithms to optimize resources. A novel approach is to utilize DT in conjunction with edge intelligence to facilitate immersive Metaverse connectivity. By optimizing resource allocation, ultra-reliable and low-latency communication (URLLC) can be realized\cite{9816050}.
The authors in\cite{Yu2022BiDirectionalDTT} explore the bi-directional dependency of DT with MEC and the interaction of DT-assisted edge computing with DT to avoid Metaverse service disruptions. Dai \emph{et al.}\cite{Dai2022ASP} integrate lightweight DT with MEC to reduce cloud load, increase the number of service requests for MEC, and significantly reduce the time complexity of the algorithm by introducing Merkle trees.
Although there have been numerous studies on industrial DTs for handling edge selection and task offloading, these schemes often overlook the data storage limitations of the Metaverse platform. In fact, the fundamental components of the Industrial Metaverse can be simply divided into virtual service providers (VSPs) and individual users and enterprises. However, not all physical assets can be deployed on the platform to capture device status information in real-time to ensure full interoperability.
Han \emph{et al.}\cite{Han2023ADH} propose a dynamic layering of platforms through IoT devices, where VSPs maintain business profitability by increasing synchronization strength. The IoT devices are responsible for transmitting data to independent VSPs. The proposed architecture achieves optimal synchronization through optimal control theory.
Zhang \emph{et al.}\cite{Zhang2022ATM} investigate the problem of multi-base station channel resource allocation in a Metaverse DT, using an end-to-end cloud three-layer framework. They design a reverse auction mechanism to solve the resource allocation and pricing problem.
Furthermore, the authors in \cite{VanHuynh2022EdgeIU} consider the joint design of communication, computation, and storage resources to formulate the problem of minimizing latency by optimizing edge task offloading and arithmetic overhead. This improves the Quality of Experience of DT in the Metaverse.

DT facilitates the digitization of physical objects, while XR technologies provide the cognitive means for interaction between the digital and physical realms. As a result, beyond the scope of DT itself, several studies explore the synergy between DT and XR technologies in industrial contexts. For example, Yang \emph{et al.} \cite{Yang2022ExtendedRA} propose a framework for smart crane applications that integrates DT and XR to enhance program development efficiency and the usability of Human-Computer Interaction (HCI) systems. Coupry \emph{et al.} \cite{Coupry2021BIMBasedDT} explore the use of XR technologies to improve Building Information Modeling (BIM) and highlight the challenges associated with implementing BIM-based systems. Jerov \emph{et al.} \cite{Jerov2020DigitalTI} examine the potential of combining DT and XR in control systems, where they develop a DT-based XR system with fundamental cognitive interaction mechanisms, linking it to real-world production scenarios. While these studies are often confined to conceptual designs or specific use cases, there is a growing need for a collaborative approach to build integrated systems that effectively combine DT and XR.

Tu \emph{et al.}\cite{Tu2023TwinXRMF} propose a TwinXR approach that effectively merges information management DT with knowledge-based XR technology. By leveraging IoT devices available in the factory, XR programs can be swiftly developed and instantiated, while the development process is greatly facilitated through the use of publicly available software packages. The authors effectively demonstrate the remarkable compatibility and versatility of this approach by showcasing its successful application to cranes and robot arms in smart factories. Nonetheless, the authors duly acknowledge that the adoption of TwinXR methods can be prohibitively expensive when the machinery is stationed at a specific, static position, rendering cost a critical limiting factor in real production settings.

\subsubsection{Application Stage in Industrial Production}

In the product design stage, enterprises initially use DT to test the production environment and build virtual models, adjusting parameters to optimize production, reduce design defect rates, and lower costs while meeting predefined requirements. This process alleviates the burden of manual trial and error, improving design accuracy and efficiency. It also involves the iterative optimization of both digital and physical models, achieving seamless connectivity through data flow\cite{Zhang2022IntegrationOD}. 

In the supply chain and operations \& maintenance (O\&M) stages, DT can optimize production processes by facilitating remote equipment maintenance and supply chain management. In equipment maintenance, DT provides real-time performance data, including current status, remaining service life, and potential issues or malfunctions, enabling operators and maintenance personnel to identify and resolve problems early, thus reducing equipment downtime. 

In supply chain management, a DT of the network is created, allowing managers to monitor and optimize operations in real-time. In the post-commercialization services phase, traditional industrial services often face issues of fragmented management and lack of unified coordination, which complicates product tracking, maintenance, and support. However, an Industrial Metaverse platform integrated with DT technology can consolidate data throughout the product lifecycle, enabling manufacturers, technical service providers, and customers to access and update product status in real-time on a unified platform. This effectively addresses the issue of fragmented management and improves service efficiency and quality.

\subsubsection{Summary}
To put it succinctly, DT can facilitate the integration of cyber-physical systems throughout the entire PLC by enabling real-time bidirectional interactions and data-driven modeling between the physical and digital worlds. This provides a new perspective for addressing challenges in industrial scenarios\cite{Onaji2022DigitalTI}. By combining explicit and tacit, structured and unstructured knowledge, real-time operational data can be layered over traditional industrial models to construct industrial DTs, helping operators better understand and manage manufacturing processes. However, existing research must build upon current DT technologies to address Metaverse-specific constraints\cite{Mourtzis2023DigitalTI}, such as security limitations and storage restrictions. Additionally, attention should be given to deployment strategies and integration with other enabling technologies, as research in this area remains at a relatively early stage.

\subsection{5G/6G Communication Network}\label{2D}
The Industrial Metaverse is typically data-driven and sensitive to perception delays in the physical layer. However, sensor data in industrial scenarios is often highly complex, with characteristics of high volume, multi-scale, and high noise. Additionally, the supporting technologies involved in the Industrial Metaverse often have high communication bandwidth requirements, such as the substantial bandwidth consumption needed for digital asset verification in NFTs, and the bandwidth demand for verifying digital tags and transactions between different entities in blockchain. Furthermore, VR systems require a data rate of 1Gbps to provide acceptable video streaming, and real-time feedback for remote operations necessitates high reliability and high-speed data connections.

Next-generation wireless communication technologies, such as 5G/6G, can provide high throughput and low latency guarantees for the upper and lower links of interactions between the physical and virtual worlds, enabling efficient management through fast transaction processing. In addition to throughput, the Industrial Metaverse aims to support high concurrency of devices within spatial and temporal bounds, ensuring that all devices and users within the platform can access and exit at any time and from any location, which increases the demand for seamless and full-coverage connectivity provided by 5G/6G. Specifically, 5G/6G technologies already support key capabilities such as URLLC, massive machine-type communication (mMTC), enhanced mobile broadband (eMBB), and network slicing, which are critical for meeting the real-time requirements of industrial services. Additionally, the new-generation wireless communication network can provide auxiliary connectivity through IoT devices for global coverage, and its inherent security technologies can also be used to defend against unknown security threats\cite{Goyal2024AnIF}.

\begin{figure}[!t]
\center{\includegraphics[width=0.5\textwidth]{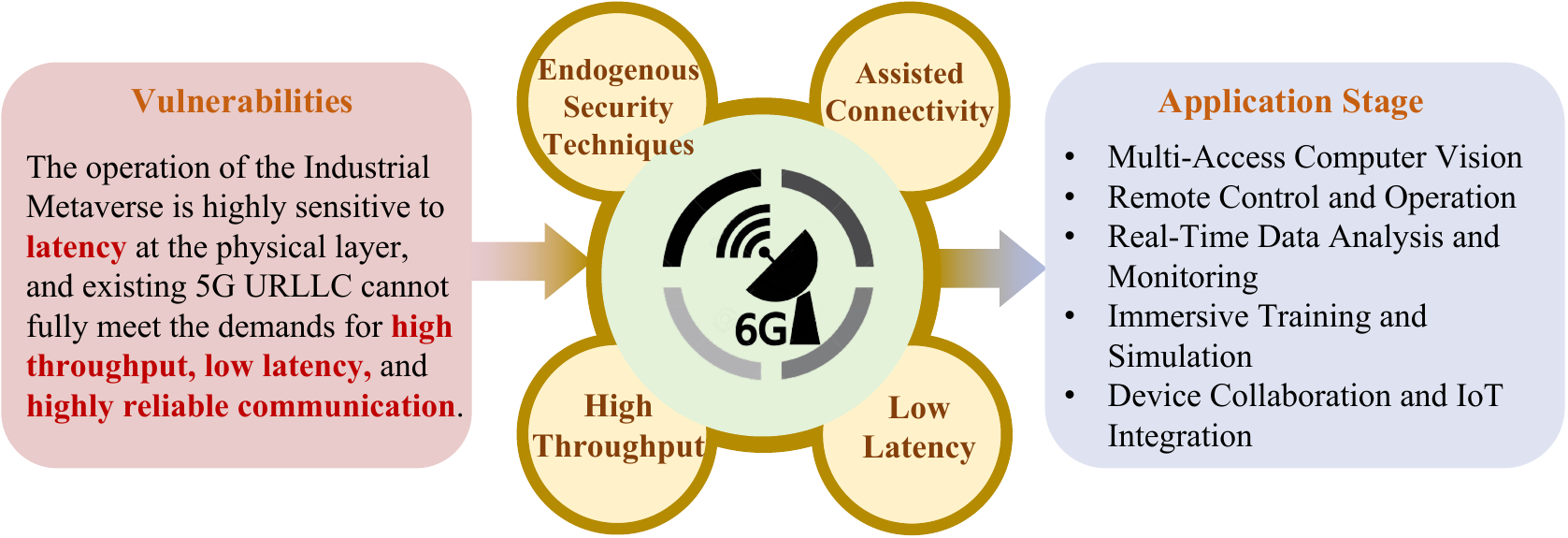}}
\caption{The vulnerabilities, advantages, and application stage of 6G in Industrial Metaverse.}
\label{fig6G}
\end{figure}

Fig. \ref{fig6G} shows the vulnerabilities, advantages, and application stage of 5G/6G in Industrial Metaverse.

\subsubsection{Research Advances of 5G/6G Enabling Industrial Metaverse}
Currently, the latest version of 5G technology is already capable of supporting XR applications. Devaki \emph{et al.}\cite{10093870} point out that fundamental XR and VR support will be enabled by 5G-Advanced (5G-A) networks; however, a complete ``Metaverse" experience cannot be envisioned until 6G is deployed. Both 5G-A and 6G offer time services with many dedicated installation-based use cases, such as wireless time-sensitive networking in the Industrial IoT and critical in-vehicle networks\cite{Morato2024ATT}. Most existing works focus on optimizing 5G/6G networks and their integration with the Metaverse, but the emphasis on different performance metrics varies by application scenario. In the Industrial Metaverse, key performance indicators include throughput, latency, reliability, age of information (AOI), and energy efficiency\cite{Huang2022AoIawareEC}. While latency and reliability are deeply coupled, the weight of other metrics also influences the joint architectural design.
In this regard, Cao \emph{et al.}\cite{Cao2023TowardIM} discuss several metrics' coupling relationship in-depth and argues that communication delay, reliability, and AoI are the three most crucial metrics in an IIoT environment with heterogeneous traffic. The authors collect data and control heterogeneous traffic from the short packet transmission optimization perspective. However, their specific task division is not given, and transmission jitter's performance loss is not negligible in the actual industrial environment.

Another approach is based on the design of an integrated architecture. MEC, as a crucial auxiliary mechanism in 5G/6G networks, can significantly reduce request latency due to its proximity to the source. In a factory setting, devices can be organized into edge computing hierarchies, including sensors, IIoT devices, and servers\cite{Li2024AoIAwareDT}. Sensors collect industrial data, IIoT devices receive and process various states, and data is then aggregated for centralized processing in the cloud. A typical scenario involves delegating lower-tier industrial operations to lightweight mobile devices, controlled remotely by administrators. Various 5G/6G mobile edge technologies are already being applied to the Metaverse, including wireless communication technologies (e.g., heterogeneous radios, intelligent reflective surfaces, non-orthogonal multiple access) and new communication technologies (e.g., semantic communication (SemCom), holographic communication, tactile communication)\cite{Bandara2024GenerativeAIwithCM}.
Muranaka \emph{et al.}\cite{10271886} propose a worker navigation system using 5G and MEC that provides services without being affected by existing industrial network infrastructure. By leveraging local 5G and MEC connections, the system can effectively avoid disruptions to the existing infrastructure.
Qian \emph{et al.}\cite{10539076} focus on optimizing user connectivity and server computing frequency allocation in an NFT-centric, BC-enabled Metaverse. By introducing the trust cost ratio as a metric, they ensured sustained user engagement and trust.
Aslam \emph{et al.}\cite{Aslam2023MetaverseF6} present a layered Metaverse design from the perspective of layers, with different layers connected by a data exchange layer. They then demonstrate the autonomous driving remote control case and finally discuss the challenges and propose solutions for 5G/6G and Metaverse integration, including how to deal with wireless communication security threats, submillimeter beamforming loss, and channel interference.
Since 6G mobile authorized devices in the same production environment need to share computational memory and frequencies, unreasonable resource allocation can easily lead to network congestion, which we will cover in detail in the next section.

In Table \ref{table3}, we compare 5G and 6G in the context of the Metaverse across multiple dimensions. In summary, both technologies provide critical network support for the Metaverse. 5G, with its current maturity and widespread adoption, lays the foundation for the initial development of the Metaverse. Meanwhile, 6G, with its higher bandwidth, lower latency, and enhanced security and computing capabilities, is poised to become a more promising support technology in the future, delivering seamless, real-time, and highly efficient interactive experiences within the Metaverse.

\begin{table*}[!t]
    \centering
    \caption{Comparison of 5G and 6G in Metaverse Applications}
    \label{table3}
    \renewcommand\arraystretch{1.2}
    \begin{tabular}{|m{4.3cm}|m{6.2cm}|m{6.3cm}|}
        \hline
        \textbf{Dimension} & \textbf{5G} & \textbf{6G} \\ \hline
        Peak Data Rate & Up to 20 Gb/s & 100 Gb/s - 1 Tb/s \\ \hline
        Latency & 10 ms & Less than 1 ms \\ \hline
        Network Density & Supports up to 1 million devices per km\textsuperscript{2} & Supports up to 10-100 million devices per km\textsuperscript{2} \\ \hline
        Bandwidth & Up to 1 GHz & Up to 100 GHz \\ \hline
        Time Synchronization Capabilities & Advanced but limited for real-time DT applications & Enhanced for precise time-sensitive applications, including XR-driven DTs \\ \hline
        Security Features & Cryptographic algorithms & Quantum-safe cryptography, enhanced security protocols \\ \hline
        Application Areas & VR/AR, IoT, immersive gaming, telemedicine & 3D holography, tactile internet, XR, AI-driven DTs \\ \hline
        Computing Paradigm & Primarily cloud computing & Edge computing with integration of quantum computing capabilities \\ \hline
        Energy Efficiency & Significant improvement compared to 4G & 10x improvement compared to 5G \\ \hline
        Integration with IoT and IIoT & Limited edge capability for large-scale IoT integration & Seamless integration with IIoT through high concurrency and low latency \\ \hline
    \end{tabular}
\end{table*}

\subsubsection{Application Stage in Industrial Production}
In multi-access computer vision scenarios, industrial IoT devices supported by 5G/6G can achieve seamless connectivity and communication, with dedicated frequencies and high data transfer rates that significantly enhance image quality within the Metaverse. This enables operators to access visual information more quickly and accurately. Furthermore, heterogeneous devices within enterprises can leverage multi-access edge computing resources over virtual 5G/6G networks to process data using deep learning technologies. This greatly simplifies operators' workloads and reduces data analysis latency and complexity. 

In the product decision and service phase, 5G/6G facilitates compute-centric communication by dynamically allocating communication resources to ensure the required Quality of Service (QoS) and achieve precise computation. By enhancing quality control, decision-making efficiency is improved, and further automation is enabled. Additionally, 5G/6G can reduce the complexity of supply chain networks, thus supporting agile and adaptive service delivery for Metaverse applications. Operators can respond and adjust more promptly in dynamically changing environments.

\subsubsection{Summary}
Firstly, in terms of time synchronization, although significant advancements have been made in 5G's time synchronization capabilities, the scope of time services in 6G still needs to be enhanced. For example, using XR to enable real-time replication of the physical world in DTs requires precise time synchronization to drive data-based decision-making for additional AI use cases, such as predictive maintenance. Accurate mirroring of the physical environment depends critically on this capability.
Additionally, due to the significant network heterogeneity among various Metaverse sub-platforms of manufacturers, Metaverse communication networks represented by 5G/6G require unified network access and control protocols. Factories deploying 5G/6G devices must also ensure the operability of existing network function devices, with implementation tailored to each individual subsystem. At present, a fully deployable 5G/6G-enabled IIoT architecture for Industrial Metaverse platforms has yet to emerge, and corresponding standards and specifications remain lacking. While some recent works have proposed related reference architectures\cite{Padhi20216GEI}, they often feature overly simplified hierarchical structures and lack empirical validation, leaving theoretical aspects unexplored.

\subsection{Extended Reality}\label{2E}
In manufacturing scenarios, there are instances wherein it becomes imperative to employ diminutive sensors or IoT devices at the distant terminus, which typically lack interfaces conducive to Metaverse user engagement.
XR, including VR, AR, and mixed reality (MR), serves as the interface between the tangible realm and the ethereal domain of the Metaverse, endowing users with a myriad of multifaceted interaction modalities\cite{Jalo2022ExtendedRT}. Within this realm, VR unveils itself as a veritable medium for sensorial emulation, employing head-mounted apparatuses to engender immersive experiences. It facilitates the visualization of diverse design iterations, verifies procedural efficacy, and bestows captivating verisimilitude. On the other hand, AR seamlessly integrates virtual components, such as textual and pictorial elements, into the fabric of reality through handheld devices like smartphones and tablets. By availing themselves of this technology, operators are endowed with instantaneous access to real-time equipment status updates, virtual cues, and instructional guidelines. Furthermore, MR transcends conventional overlays of virtual content upon tangible surroundings, instead forging a harmonious convergence between the virtual and the corporeal, facilitated by perceptive and interpretative faculties attuned to the veritable realm.

\begin{figure}[!t]
\center{\includegraphics[width=0.5\textwidth]{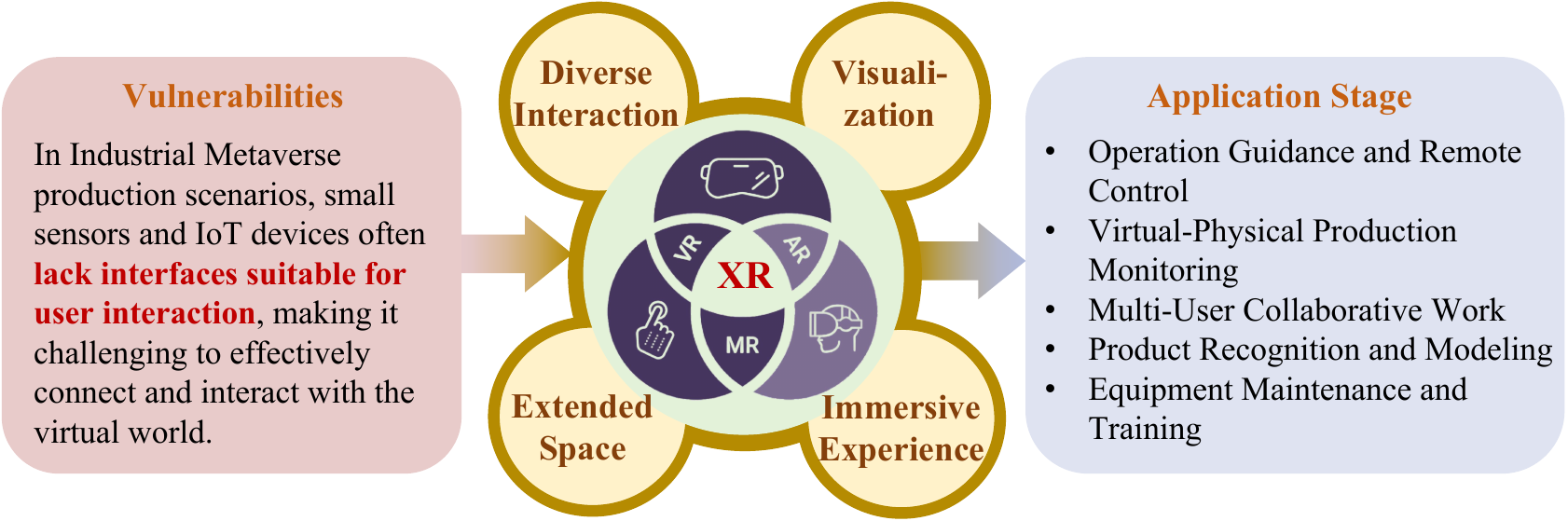}}
\caption{The vulnerabilities, advantages, and application stage of XR in Industrial Metaverse.}
\label{figXR}
\end{figure}

In broad terms, constrained by diverse factors in industrial scenarios, comparatively substantial apparatuses like robotic arms, for instance, manifest greater aptitude for remote manipulation by technicians, whereas XR can function as an ad hoc controller. Furthermore, XR technology has the capability to furnish a fusion of tangible and virtual domains for user-machine interplay, thereby lending itself to deployment within industrial production scenarios for the purposes of operational guidance, equipment maintenance training, and product modeling and design.
Fig. \ref{figXR} shows the vulnerabilities, advantages, and application stage of XR in Industrial Metaverse.

\subsubsection{Research Advances of XR Enabling Industrial Metaverse}
In a certain context, XR constructs an adaptable and fluid realm for visualizing virtual content within the realm of MR. However, a primary concern arises regarding the imperative of upholding the sustainable connectivity attributes of content within the physical realm, as virtual contrivances in the Industrial Metaverse persistently expand. Should an information disjunction arise between the Metaverse and the physical space, coupled with a dearth of consistent communication, it can potentially engender cognitive overload for the user\cite{Guan2022ExtendedRA}. Put simply, when operators engage in interactions within the virtual plant, they are typically unable to access and manipulate the genuine physical apparatus. Consequently, it is crucial to address these disconnections within the Metaverse in a scientifically comprehensive manner, as failure to do so can easily impinge upon critical industrial scenarios.

An intriguing concept in this realm is the utilization of an extended Metaverse agent approach based on the MiRAs cube theory\cite{Holz2011MiRAM}. This approach serves the purpose of dynamically governing virtual device control and hologram modeling, while facilitating interactions with other holograms and operators to manifest within the physical realm. Building upon this notion, Rincon \emph{et al.}\cite{Rincon2016ExtendingMM} leverage agents founded on the meta-model (MAM5) and JaCalIVE frameworks to mitigate discrepancies between simulated and real systems. Intelligent resource artifacts are incorporated, enabling seamless access to the physical world for devices. The dynamic essence of agents is further explored by Croatti \emph{et al.}\cite{Croatti2018AMA}, whose architecture allows for agents to be dynamically altered at runtime, employing diverse heterogeneous agent programming techniques to modify agents within AR. In addition to system modeling, certain researchers have embraced agent-based approaches to automate XR quality testing and support experience evaluation by adopting user roles\cite{Prada2020AgentbasedTO}. Braud \emph{et al.}\cite{Braud2022DiOSAnER} construct an XR architecture from the operating system level, which perceives the physical world as a shared resource among applications. By implementing meticulous control mechanisms through several components such as environment understanding, specialized chip drivers, networking, interoperability, and various other interconnected elements, they achieve enhanced integration of content with the physical-digital world, thereby attaining superior performance.
Drawing upon the extended Metaverse architecture, Guan \emph{et al.}\cite{Guan2022ExtendedRA} conduct an in-depth analysis of interface design consistency within the Metaverse. They attribute inconsistencies to signal noise and propose the utilization of ``noise reduction" techniques to mitigate disparities between the Metaverse and the physical world. This approach aids in facilitating the development of interoperable interfaces. The authors further delve into the realm of seamless integration by exploring the presence of interoperable agents. They showcase the effectiveness of this approach through two early design projects, presenting interactive and tangible Metaverse applications that can be deployed across both physical and virtual realities.
As interfaces of this nature continue to be researched and developed, the integration of suitable filters and controls becomes crucial in ensuring the sustained continuity of the Extended Metaverse. This integration holds the potential to serve as a valuable research direction in the long-term future.

Another intriguing concept in interactive device interfaces is the utilization of a convergent technological paradigm known as XR-IoT (XRI) systems, which combine XR and IoT. These XRI systems capitalize on the complementary attributes of both domains, leveraging the interconnectivity of environmental entities facilitated by IoT, while XR empowers information-rich interactive interfaces and serves as a platform for HCI\cite{Shao2019IoTAM}.
The evaluation of interface design in XRI systems is initially proposed by Dnser \emph{et al.}\cite{Dnser2011EvaluatingAR}, who delineat the original design intent of XRI and presented a comprehensive evaluation framework from the perspective of interface design, with usability at its core. Building upon this work, Tsang \emph{et al.}\cite{Tsang2021AHQ} provide a more exhaustive taxonomic approach and a QoE reference standard for the development and assessment of XRI systems.
In the context of real-world industrial scenarios, Morris \emph{et al.}\cite{Morris2021AnXM} envision an immersive, multi-user, and multi-agent-driven XRI-based intelligent workstation system. This system incorporates haptic feedback-equipped desks, cameras, and other devices to monitor personnel status, while physical laptops run machine learning models to determine personnel conditions.
Taking it a step further, Oppermann \emph{et al.}\cite{Oppermann2023IndustrialMS} introduce a 5G XR toolbox tailored for remote maintenance scenarios in industrial equipment. This toolbox facilitates MR and VR views by establishing digital connections between field equipment and personnel. Operating on actual machines within real production lines, the system utilizes authentic CAD data to create a realistic prototype environment.

Another intriguing aspect to explore is whether the use of AR/VR devices could have adverse effects on workload, as workload primarily reflects the input costs in production tasks. Kantowitz \emph{et al.} \cite{Kantowitz2020MentalW} demonstrate that efficiency tends to decrease as task complexity increases during HCI, even within the limits of maximum workload capacity. Simply put, this raises the question of whether operators need to exert additional cognitive effort to perform tasks through VR interfaces that could otherwise be completed in the real world. Furthermore, Zhao \emph{et al.} \cite{Zhao2022OnTP} highlight that by creating multisensory experiences such as visual, auditory, and tactile sensations in XR immersive environments, learners' emotional engagement and cognitive levels can be effectively enhanced within the Metaverse. Future research could explore the mechanisms and boundary conditions of these various dimensions, providing XR designers with a more scientifically informed approach to developing HCI systems.

\subsubsection{Application Stage in Industrial Production}
In product recognition and modeling, object detection is widely applied in XR (Extended Reality) and is essential for realizing the Metaverse. For example, systems can leverage computer vision technologies for real-time object recognition and overlay relevant information on displays, such as product specifications, usage instructions, and maintenance methods. This technology can also be applied to industrial product modeling and visualization. Real-time visual feedback allows operators to intuitively understand and adjust products, reducing design and assembly errors and improving the efficiency of product design, assembly, and maintenance. 

In device pose recognition, XR environments often require the observation and recognition of device movements, such as robots, within a 3D immersive factory setting, and the generation of feedback based on specific actions. For instance, in assembly lines, AR technology can detect the pose of devices or components in real-time, including rotation angles, positions, and dimensions, thus assisting operators in accurate assembly and adjustments.

\subsubsection{Summary}
The Industrial Metaverse is a collaborative virtual space that enables enterprise employees to interact and work together in a digital environment. By manifesting as tangible virtual images within this space, operators can engage in virtual production and collaboration that runs parallel to the real world, thereby enabling digital transformation. This immersive technology is poised to shape the new form of the Industrial Internet. Additionally, AR and MR technologies have the potential to revolutionize the operation of the physical world within factories. As digital entities transition from a purely virtual environment to a physical one, further industrial design and technological considerations are necessary to ensure the success of digital transformation.

Here's the table without the blue font color commands:

\begin{table*}[!t]
\begin{center}   
\caption{Research on enabling technologies in the Industrial Metaverse.} 
\label{table4}
\renewcommand\arraystretch{1.12}
\begin{tabular}{|m{0.6cm}<{\centering}|m{0.5cm}<{\centering}|m{0.5cm}<{\centering}|p{1.8cm}|p{3.4cm}|p{6cm}|p{2.5cm}|}    
\hline     
\textbf{Tech.} & \textbf{Ref.} & \textbf{Year} & \textbf{Scenes} & \textbf{Objectives} & \textbf{Methods}&\textbf{Limitations} \\  
\hline      
BC & \cite{Lin2022TowardsMM}
& 2022
& Data collection to storage.
&  Ensure credible execution of Industrial Metaverse.
& A BC-based trusted collaborative governance system for Metaverse smart manufacturing scenarios.
& Lack of simulation.\\
\cline{2-7}
& \cite{Kang2022BlockchainbasedFL}
& 2022
& Cross-enterprise collaboration.
& Improving privacy protection in the Industrial Metaverse.
& Combining joint learning and cross-chain techniques to design a hierarchical BC architecture with main chain and sub-chains.
& Lack of mechanisms to prevent malicious local devices.\\
\cline{2-7}
& \cite{Xu2022MetaverseNC}
& 2022
& Access management.
& Responding to Metaverse and decentralized Internet efforts.
& Dedicated frequency bands in regional spectral bands for link transmission.
& Lack of simulation.\\
\cline{2-7}
& \cite{Wang2022BlockchainbasedER}
& 2022
& Cross-enterprise collaboration.
& Optimizing resource sharing and utilization.
& A practical BC-based MEC platform is proposed for resource sharing and optimal utilization.
& No detailed assessment is provided.\\
\hline
PPC 
& \cite{lyu-etal-2020-differentially}
& 2020
& General.
& Reducing privacy leakage without loss of performance.
& DPNR with differential privacy for formal privacy guarantee.
& Introduce noise in complex tasks.\\
\cline{2-7}
& \cite{Liu2023WassersteinGA}
& 2023
& Data sharing.
& Balancing privacy and utility in shared data.
& WGAN for privacy-preserving data sharing.
& Real-world deployment.\\
\cline{2-7}
& \cite{Sun2024FedKCPF}
& 2024
& Healthcare services.
&  Ensure privacy against model poisoning attacks.
& A personalized FL algorithm.
& Scalability challenges.\\
\cline{2-7}
& \cite{Aristodemou2024BayesianOP}
& 2024
& Cross-enterprise collaboration.
& Assess PFL's vulnerability to poisoning attacks.
& Bayesian optimization-driven attack.
& Limited real-world applicability.\\
\hline
DT & \cite{VanHuynh2022EdgeIU}
& 2022
& Cross-enterprise collaboration.
& Joint design of communication, computing and storage resources.
& The latency minimization problem is formulated by optimizing edge task offloading, arithmetic overhead, and thus improving the QoE of DT in the Metaverse.
& The bias is assumed to be fixed.\\
\cline{2-7}
& \cite{Zhang2022ATM}
& 2022
& Base station data receiving.
& Ensuring the quality of VSP's services.
& The authors study resource allocation in a Metaverse DT with a cloud framework.
& Storage resources are not considered.\\
\cline{2-7}
& \cite{Wu2022DigitalTO}
& 2022
& Surface defect detection.
& DT solutions for small surface defect inspection tasks.
& Integration of multimodal data.
& Algorithms struggle with large data.\\
\cline{2-7}
& \cite{Dai2022ASP}
& 2022
& Base station data receiving.
& Reduce cloud load with MEC.
& A Service Placement Algorithm Based on Merkle Tree.
& Baseline algorithm is not enough.\\
\cline{2-7}
& \cite{Yang2022ExtendedRA}
& 2022
& Crane application.
& Improved efficiency of XR application development.
& The architecture includes a perception layer, ROS model, and Unity service layer for XR apps.
& It may not apply to other areas.\\
\cline{2-7}
& \cite{Tu2023TwinXRMF}
& 2023
& Supply chain and O\&M.
& Promoting synergy between DT and XR.
& TwinXR combines information management DT and knowledge-based XR.
& Feature mapping is entirely dependent on shared data structure.\\
\hline
5G/6G & \cite{DBLP:journals/jcin/ChangZLXGSXKNQW22} 
& 2022
& Computer vision.
& Organic integration of edge AI and Metaverse with 6G support.
& Three novel edge Metaverse architectures are introduced that use 6G-enabled edge AI to address resource and computational limitations.
& Missing network latency and privacy issues with 6G.\\
\cline{2-7}       
& \cite{Chua2022ResourceAF} 
& 2022
& Product decision and service phase.
& Allows Metaverse mobile users to download virtual world scenes in real time.
& An environment with multiple cellular stations is designed where the task of downloading graphics would be handed over between cellular stations.
& The role of AI for 6G is not addressed.\\
\hline      
XR & \cite{Croatti2018AMA} 
& 2018
& Hologram modeling.
& Developing agent-based AR/MR systems.
& Support for modifying agents in AR using different heterogeneous agent programming techniques.
& Lack of integration with standards.\\

\cline{2-7}      
& \cite{Warin2022VisionUP} 
& 2022
& General.
& Understanding XR's impact on user privacy.
& The article analyzes the gap between user privacy perceptions and their specific behaviors across XR.
& Omission of audit and transparency.\\
\cline{2-7}      
& \cite{Braud2022DiOSAnER} 
& 2022
& Automated quality tests.
& Operating system level XR architecture.
& Get better performance by enforcing fine-grained control.
& No mention of actual use cases for XROS.\\
\cline{2-7}       
& \cite{Xi2022TheCO} 
& 2022
& Workload metrics.
& Correlation of AR to overall workload.
& NASA-TLX for measuring subjective workloads.
& Subjective experiment.\\
\hline      
AI 
& \cite{Wong2022GetWildAV} 
& 2022
& Product design.
& Boost 3D modeling efficiency.
& AI-assisted VR modeling.
& Lack of baseline.\\
\cline{2-7}       
& \cite{Yang2022ThreeDimensionalSN} 
& 2022
& Product design.
& Memory interfaces have limited bandwidth.
& Higher XR resolution can be achieved by stacking multiple layers of memory
 using 3D stacking.
& No practical deployment examples.\\
\cline{2-7}       
& \cite{Badruddoja2022TrustedAW} 
& 2022
& Digit recognition.
& Mitigating security risks in the Metaverse.
& Arming smart contracts with intelligence and using AI algorithms for prediction.
& Missing energy overhead.\\
\cline{2-7}       
& \cite{Khalaj2022MetaverseAA} 
& 2022
& Product analysis.
& Improving the reliability of numerical modeling.
& Four ML methods create advanced DTs for 42SiCr alloy, accounting for uncertainties and nonlinearities.
& The industrial scenarios are homogenous.\\

\hline   
\end{tabular}   
\end{center}  
\end{table*}

\subsection{Artificial Intelligence}\label{2F}
As the foundational cornerstone of the Industrial Metaverse, AI obtains its input data through an array of sensor devices situated within the factory. This data, once subjected to meticulous preprocessing procedures, serves a multitude of industrial production processes, encompassing the realms of image and audio recognition, astute decision-making support, and production planning, among others\cite{DBLP:journals/corr/abs-2202-09027}. Concurrently, AI remains attuned to user interactions transpiring within the virtual realm. Furthermore, beyond its inherent industrial applicative worth, AI possesses the potential to harmoniously unite with other pivotal enabling technologies, thereby bestowing an enhanced HCI experience. All in all, the pivotal role of AI technology lies in its provision of essential implementation reliability and exceptional performance standards to propel the Metaverse forward.

\subsubsection{Research Advances of Artificial Intelligence Enabling Industrial Metaverse}

At this juncture, the research endeavors surrounding AI in the Industrial Metaverse converge upon two primary focal points. Firstly, the synergy between AI and other enabling technologies takes center stage, encompassing algorithm optimization for VR devices, as well as deployment paradigms stemming from the amalgamation of BC and MEC. Such combinations give rise to distributed AI and edge intelligence, which find practical application within the Industrial Metaverse. Secondly, dedicated technical research pertaining to industrial AI itself unfolds, including domains such as natural language processing, speech recognition, and the notable emergence of generative AI models, typified by chatGPT, which exhibit broad applicability in domains like intelligent Q\&A and assisted decision-making.

First, the AI-assisted XR mechanism includes technologies such as computer vision and XR computation, enabling real-time capture of environmental data within the field of view and delivering immediate processing results to users via display devices. In terms of 3D modeling, beyond basic model generation and editing, AI technology can be utilized for automated model optimization and the generation of complex geometric structures. For example, Generative Adversarial Networks (GANs) and Variational Autoencoders (VAEs) are employed to automatically generate new designs, enhancing flexibility and customization capabilities in design. On one hand, research into adapting architectures for XR devices has been conducted, with Wong \emph{et al.}\cite{Wong2022GetWildAV} developing a system named GetWild. The core network at the AI stage of this system is a CGAN, which allows users to capture images using VR devices as input to create rough 3D objects and terrain, followed by further editing operations. This technology holds practical applications in industrial product design. However, the extensive 3D modeling can pose resource limitations for small-scale edge devices like VR headsets. Even with workload optimizers tailored for XR devices, high memory requirements still result in latency issues. To address this, Yang \emph{et al.}\cite{Yang2022ThreeDimensionalSN} utilize a 3D-stacked neural network accelerator architecture to evaluate energy and latency improvements. This simulator estimates various hardware metrics, such as execution time (latency), energy consumption, input/output traffic, and resource utilization of prototype ML accelerators running AR/VR models. Based on analysis models, the simulator extracts operators and calculates expected performance metrics using energy tables based on accelerator architecture, guiding architectural design decisions for different ML models. Subsequent work can further expand memory capacity along the X, Y, and Z dimensions to identify points of diminishing returns and consider the practical deployment and implementation of platforms.
Regarding VR performance estimation, Yang \emph{et al.}\cite{Yang20183DPV} also provide a dataset for VR quality assessment (VRQ-TJU) and introduced an end-to-end 3D CNN for video quality evaluation. The 3D CNN takes VR differential video patches as input and considers information across different frames. Future work can analyze multiple characteristics, not limited to spatial distribution changes. In addition to architecture optimization and quality assessment, AI is widely used for VR image denoising, super-resolution generation, and more, to offer finer operational granularity and clearer interaction interfaces, supporting the intricate needs of industrial applications\cite{Yeh2019MultiScaleDR}.

Another trending direction entails AI-assisted biometrics for human pose recognition in complex environments and user device verification within the Metaverse's realm. The former encompasses motion recognition and body part tracking facilitated through the synergy of controllers and motion sensing devices. Such applications commonly manifest in player interaction scenarios within gaming contexts, but fall outside the purview of this discourse. The latter, primarily rooted in user privacy protection, will be expounded upon in the ensuing section.

In addition to integration with XR, a Metaverse architecture supported by BC, Web 3.0, and MEC has further spurred the development of decentralized AI (DeAI)\cite{Liu2020PathPF}. Centralized AI (CeAI), as a static and centralized system, suffers from low flexibility and poor scalability, making it unsuitable for cross-enterprise collaborative scenarios. On the other hand, DeAI can break down tasks issued by central departments into subtasks, which are then distributed to lower-level agents for processing, ultimately forming a collaborative final solution\cite{Wu2024KnowledgeEnhancedDG}. For example, Li \emph{et al.}\cite{Lin2024BlockchainBasedEA} design a verification mechanism based on smart contracts and a content caching mechanism using Stackelberg games to prevent random outcomes from AI Generated Content (AIGC) services, maximizing the benefits of Metaverse participants.
While intricate AI models often entail substantial computational overhead, rendering real-time execution on edge devices within the Metaverse a challenge, lightweight AI processing solutions have matured. These encompass techniques such as parameter pruning, quantization, low-precision approximation, and knowledge distillation. 
Lee \emph{et al.}\cite{Lee2019PredictingTT} have successfully applied a lightweight LSTM model in Head-mounted Displays (HMDs). The network takes a set of 10 data points sampled from motion data at intervals of 5 timestamps as input. Subsequently, online learning techniques can be applied for personalization to enhance accuracy.
By virtue of lightweight edge deployment, DeAI enables the realization of intelligent device agents at the edge level. Nodes within the edge layer undertake preliminary solutions through local reasoning and subsequently coordinate and communicate with neighboring nodes, thereby facilitating resource sharing. This paradigm presently represents the optimal approach for the Industrial Metaverse.

Another topic concerning AI enabling technologies centers around AIGC. With its advanced capabilities in digital content bundling, intelligent content editing, and smart content creation, AIGC technology is rapidly becoming the foundation for content creation and delivery, and is regarded as the ``engine" driving the Metaverse\cite{Wang2024ExploringAI}. Consider an employee equipped with an AR device tasked with arranging equipment in a virtual environment. In this scenario, the employee may require a catalog of pre-ordered equipment, prompting the input of relevant details into an AIGC model. The model then generates equipment entities based on the given context and scenario. Notably, AIGC within the Metaverse should meet two essential criteria: first, it should support device access at any time and any location; second, it should maximize user utility within the Metaverse while satisfying user needs. Nevertheless, due to variations in task suitability among AIGCs and the non-uniform distribution of computing power across edge servers, a resource scheduling scheme that ensures reasonable allocation becomes imperative. 
Addressing this concern, Zeng \emph{et al.}\cite{Zeng2024DelayAwarePO} consider multimodal parallel offloading scenarios, minimizing latency in wireless environments during multimodal AIGC service parallel offloading by using multimodal data RP segmentation and parallel load migration. Lin \emph{et al.}\cite{Lin2023AUF} explore the joint deployment architecture of SemCom and AIGC, including a stepwise workflow to capture integration and coordination benefits, concluding with a discussion of several potential use cases.
In summary, the emergence of AIGC has brought new promise to the development of the Metaverse, such as diversifying content and reducing repetitive production labor. Additionally, Chen \emph{et al.}\cite{Chen2024AIGCbasedED} highlight that AIGC can complement other enabling technologies. For instance, integrating AIGC into DT networks can help explore network behaviors such as routing, resource allocation, and network fault diagnosis\cite{10301748}. Conversely, DT can enhance AIGC performance by optimizing data collection strategies, such as node deployment strategies, sampling frequency, and incentive mechanisms, thus improving data collection efficiency and alleviating bandwidth pressure through physical environment simulations. Of course, future work should not only continue to develop the advantages and potential of AIGC but also pay attention to regulating generated content and filtering explosive content\cite{Lee2023WhatIW}.

AI is poised to enable intelligent networking, immersive digital worlds, inclusive user interfaces, accurate avatar creation, multilingual accessibility, and many other features within the concept of the Metaverse, which is inherently user-centric\cite{Wiederhold2023TreadingCI}. The more precise and refined the AI models are, the better the overall experience can be for users. Multilingual accessibility and automatic translation are also critical components that facilitate cross-border enterprise access to the Metaverse. Nevertheless, interpretable AI is essential to enhancing brain mapping accuracy, as it enables explanations for trained models and enhances the interpretability of AI black box operations. In recent years, a variety of interpretable AI-based approaches have been introduced to improve control and monitoring of unwanted or adverse effects, such as biased decision making. Interpretable AI utilizes semantic representations to uncover relationships between different object coverages in images, thereby enhancing the interpretability of deep learning solutions. Additionally, interpretable AI can be utilized for product design and marketing, helping to analyze competitor performance and identify opportunities to outperform them\cite{Han2021ExplainableAI}. By leveraging customer-generated reviews, interpretable AI can extract competitive factors for products and effectively reflect customer opinions.

\subsubsection{Summary}
AI Technologies in the Industrial Metaverse typically need to meet requirements such as real-time performance\cite{Xu2022AFD}, scalability\cite{Zhai2022EducationMI}, and interpretability\cite{Bibri2022TheMA}. Among these, the aspect of interpretability still presents a significant gap in research. In many AI-assisted services and applications within the Metaverse, decisions are made by AI agents, which are often driven by machine learning (ML) models functioning as black boxes, lacking the ability to explain their decisions. As a result, developers, virtual world designers, and users in the Industrial Metaverse may find it difficult to understand the mechanisms behind AI-driven decisions. This lack of interpretability poses potential risks; even a 1\% decision-making error in industrial manufacturing can have unacceptable consequences, affecting not only economic outcomes but also the safety of production processes.

By developing more transparent AI models, operators' trust in these systems can be enhanced, enabling more effective use of AI technology to support decision-making and operations. Such improvements contribute to creating a more efficient and safer collaborative environment within the Industrial Metaverse.

\section{Existing issues in Industrial Metaverse}\label{3}
\begin{figure*}[!t]
\center{\includegraphics[width=0.68\textwidth]{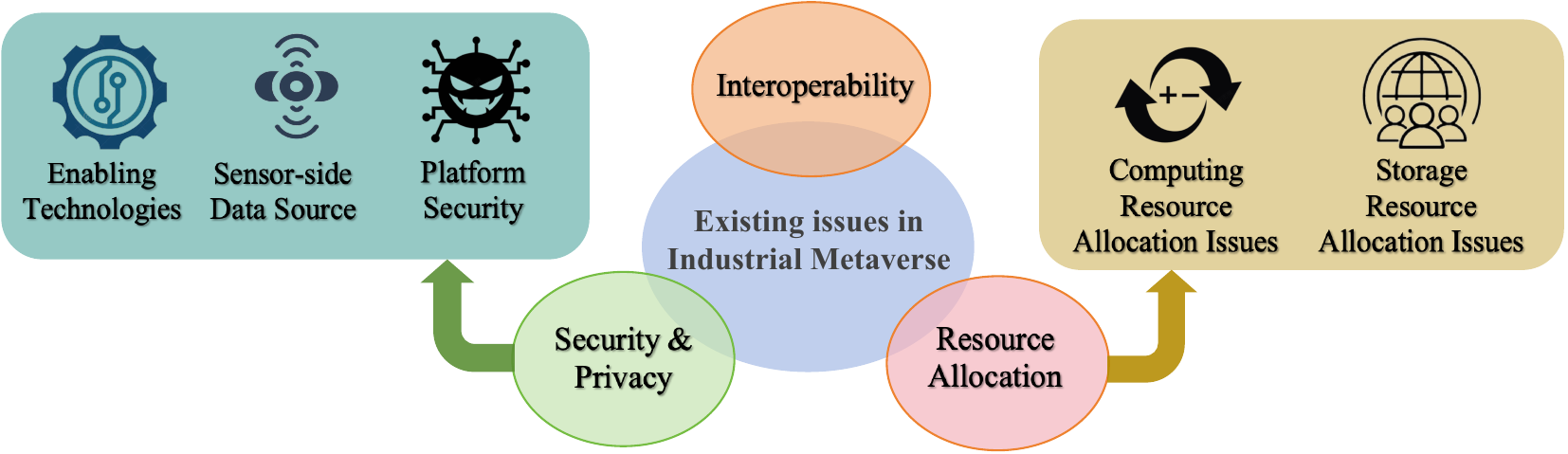}}
\caption{Existing issues in Industrial Metaverse. Including security and privacy, resource allocation, and interoperability. In terms of security, it can be discussed from the perspectives of sensors, platforms, and enabling technologies themselves. Resource allocation includes two aspects: computing resource allocation and storage resource allocation.
}
\label{fig4}
\end{figure*}

In this section, we will delve into the crucial issues that currently exist in the Industrial Metaverse. With the rising complexity of connections and interactions between systems and devices, security threats and vulnerabilities have become a more pressing concern. Furthermore, the vast number of sensors, devices, and data involved in the Industrial Metaverse require efficient data management to prevent privacy breaches during data transmission\cite{Njoku2022ProspectsAC}. Given that most enterprises prioritize profit maximization and face resource constraints, scientific resource optimization is necessary to reduce costs. We will elaborate on these challenges in the following sections.
Fig. \ref{fig4} illustrates the existing issues in Industrial Metaverse.  
Table \ref{table5} summarizes research efforts addressing existing challenges and issues in the Industrial Metaverse.

\subsection{Security and Privacy}\label{3A}
In the previous section, we outlined the immersive experiences and new management models brought by the Industrial Metaverse. However, this also introduces unprecedented security threats\cite{Ma2019OnSP}, with challenges related to security and privacy throughout the entire production cycle, from data collection to final decision-making. Specifically, the integrity and availability of industrial data (e.g., ensuring that data, from the sensor collection point to its input into the model, remains unaltered and undamaged) and the authentication, authorization, and traceability requirements for employees (i.e., ensuring that operations on devices and access to data go through a strict validation mechanism, allowing only authorized users to access them) are key issues that need to be addressed. 

While some techniques mentioned in the previous section can anonymize data and user identities, they do not solve all security problems. 
We will organize existing solutions based on the architecture shown in Fig. \ref{fig3}. Then, we will discuss the security and privacy issues that may arise at different stages, including data collection, cross-component data transmission security threats, and internal security problems within components. Finally, we will explore potential research directions.

\subsubsection{Sensor-side Industrial Data Source}

To enhance the immersion of the Metaverse and achieve more precise industrial decision models, industrial sensors play a pivotal role in collecting extensive data for the platform. These data sources encompass not only initial training samples and physical parameters used to construct the virtual space but also user information.

When it comes to industrial source data, the emphasis lies in safeguarding the data itself. For instance, redundant information in visual content, which does not contribute to the training of models but is indiscriminately collected by sensors, can be obfuscated by overlaying a mask to mitigate potential privacy breaches. Validated operators can have their virtual images in the Metaverse replaced with generic mappings. Another approach involves employing black and white list mechanisms, where only specific information is retained, and any data outside the whitelist is immediately discarded and inaccessible. Wu \emph{et al.}\cite{Wu2021DAPterPU} propose a data misuse prevention mechanism called DAPter, wherein input data undergoes lightweight generator transformations to remove unnecessary information with minimal additional overhead. The blacklisting mechanism is commonly utilized for safeguarding user facial information. In this method, faces are decomposed into vector signals, comprising identity and attribute features. Subsequently, an anonymization model is generated to integrate and reconstruct the vectors based on the need for further processing. This approach can also be employed for classifying features of industrial devices, such as device models and metric parameters. Additionally, traditional data encryption and storage schemes used in cloud-side end systems can be applied to Industrial Metaverse platforms. These include classical symmetric and asymmetric encryption techniques, cloud storage encryption, disk encryption, and more.

Apart from the issue of data leakage, the quality of data sources is also classified as a data security concern. On one hand, malicious users may generate low-quality or even false data to disrupt virtual world project construction. On the other hand, profit-oriented participants may lower data verification standards to control costs, leading to misleading data analysis and model training. The first case can be addressed through identification of malicious users and tackling data-related attacks. Common data attacks include data injection attacks and data tampering attacks. Data injection attacks involve perpetrators injecting false data into industrial systems through illicit means to mislead and disrupt system operations. Countermeasures against such attacks include implementing detection and verification mechanisms, as well as access control. For example, Pang \emph{et al.}\cite{Pang2022FalseDI} employ a Kalman anomaly detector based on residuals to securely control networked systems, while Guo \emph{et al.}\cite{Guo2022DetectionOS} use a random coding scheme to detect residuals, transforming the design problem of coded signals into a constrained nonconvex optimization problem. Access control, which restricts data modification to authorized personnel only, is widely employed in industrial settings. Existing schemes include access control architectures based on BC\cite{Bera2021AIEnabledBA}, cryptographic storage methods\cite{Ruj2014DecentralizedAC}, and more. These access control schemes can also be applied to combat data tampering attacks. As for the second case, various trustworthiness identification and processing schemes have been established to address the issue of low-quality data sources in industrial settings. These schemes encompass sensor node trustworthiness tracing\cite{Shen2021ATTDCAA}, mining association rules using data mining algorithms\cite{Palacios2012MiningFA}, methods for prognosis of low-quality data integration\cite{Duan2022AnEP}, and data enhancement techniques\cite{Yang2022Mask2DefectAP}.

\subsubsection{Data and Network Security Threats in Distributed Industrial Platforms}
The distributed architecture of the Industrial Metaverse platform is well-suited for handling large-scale data processing scenarios. However, the introduction of massive P2P links also brings about increased risks of data leakage and management challenges. On one hand, the lifecycle of the Industrial Metaverse spans across different stages, with multiple dispersed stakeholders performing various operations. The increase in digital connections may expose enterprise assets or production processes to data leakage risks. However, leveraging BC-based distributed storage can effectively manage data throughout the PLC. The storage layer provides secure distributed data storage with lightweight and scalability features. Suhail \emph{et al.}\cite{Suhail2021BlockchainBasedDT} discuss deployment schemes for BC-based Metaverse platforms. They utilized multiple sensors with overlapping perspectives for cross-validation to ensure the credibility of on-chain data. Empowering decision-making through BC in various industrial use cases addresses challenges such as different data repositories and insecure propagation. Peng \emph{et al.}\cite{Peng2023DistributedIF} further consider edge offloading in industrial cloud-edge collaborative architectures, utilizing DT to capture temporal features of computational resources. Additionally, in the second incentive phase, they integrated privacy-security investments to assist industrial agents in making offloading decisions. Shen \emph{et al.}\cite{Shen2021SecureSO} integrate distributed storage and cloud storage, encrypting raw virtual data in the cloud, storing transaction records in the BC, and maximizing overall benefits for manufacturers and buyers through optimal sampling selection. However, for industrial sensors with higher sampling rates, limitations of real-time sharing need to be further considered. 
Furthermore, BC is not a one-size-fits-all solution, as issues such as decision paralysis due to information overload, high energy consumption, and latency may arise. Moreover, the nature of the ``black box" from BC compels workers to trust the integrity and fairness of processes without understanding the technical basis. 

Another issue arising from the distributed architecture of the Industrial Metaverse is the potential for errors and attacks due to multiple untrusted but interdependent participating entities and the integration of non-intersecting data from multiple sources. For instance, man-in-the-middle attacks or altering communication between users and servers may intercept/modify data packets, compromising the confidentiality, integrity, and availability of the system. Cross-site scripting attacks inject external malicious JavaScript code into websites, enabling attackers to inject malicious JavaScript code directly into client websites or have client websites request storage locations. Immersive interaction involves many devices such as wearable devices, headsets, base stations, and controllers, with large-scale data exchange.
Despite manufacturers implementing encrypted communication, industrial data may still be accessed by adversaries through eavesdropping methods during transmission. Currently, efforts are underway to enhance industrial network security and anomaly detection. Kim \emph{et al.}\cite{Kim2020APADAP} utilize a system based on autoencoders and long short-term memory models to learn normal values of sensors and actuators for monitoring network attacks. Kwon \emph{et al.}\cite{Kwon2020IEEE1P} study distributed network protocols widely used in distribution and transmission systems, proposing an anomaly detection method based on bidirectional recurrent networks and IEEE 1815.1. They verified the method with attack data from networked physical systems, successfully detecting eight types of attacks, including unauthorized command attacks and false data injection. 
However, existing detection systems deployed in Metaverse platforms often result in high computational costs and delays and are difficult to adapt to dynamic attack environments (such as atypical and polymorphic attacks). Therefore, network security researchers must further optimize and promote existing strategies to establish effective models.

\subsubsection{Inherent Security Vulnerabilities of Enabling Technologies}

The enabling technologies mentioned in Section \ref{2} of the Industrial Metaverse also encompass significant security vulnerabilities that cannot be disregarded.

Concerning AI technology, although Wang \emph{et al.}\cite{Wang2022ASO} highlight its potential for detecting illicit entities and anomalous accounts within the Metaverse, such as discerning suspicious signals through correlating user-generated textual information and spatiotemporal activities, AI also poses various potential risks\cite{He2021DatingSecDM}. On one hand, the industrial platform interconnects a wide array of industrial devices, Metaverse users, collaborative enterprises, and applications, necessitating automated regulation and decision-making due to the intricacies of data management. This grants algorithms extensive operational privileges to reduce managerial costs. However, this approach may give rise to issues like biased decisions\cite{CorbettDavies2017AlgorithmicDM}, low transparency\cite{Datta2016AlgorithmicTV}, and operational fragility\cite{Cresci2021AdversarialML}. While certain existing work suggests hierarchical governance structures for algorithm accountability or assesses potential risks of algorithm engines based on AI explainability, a unified compliance governance protocol is yet to be established. On the other hand, due to the black-box nature of AI, malevolent attackers can infiltrate the enterprise source code with malicious code to cause system-level disruption. For instance, they may employ GANs to fabricate false virtual data or embed malicious software during the model training process, which developers often find challenging to detect in early training results. Therefore, in the industrial context, modular interpretable AI may serve as a promising avenue for future research.

In the application of distributed ledger technology, the Industrial Metaverse platform records user information and production data, and enables anonymous P2P interactions through encryption methods, providing reliable data storage and transmission for enterprises and users. However, this also presents security risks, some of which stem from the consensus mechanism, such as PoW. If an attacker’s computational power exceeds half of the total power in the BC, they can take control of the entire platform system. Additionally, while consensus mechanisms can provide stronger tamper-proof capabilities for industrial data, the computational overhead generated by existing mainstream consensus mechanisms is also a significant concern\cite{Jiang2024ProofofTrustedWorkAL}. Another risk arises from the key distribution mechanism, as decentralized key management means that private keys may be concentrated in an entity or server, and if that entity is attacked, all keys could be compromised. Currently, multi-signature mechanisms can be used to prevent private key leakage, and offline storage can help protect against network attacks.

In wireless communications, although 5G/6G wireless systems supported by the Metaverse bring many advantages, they also face security challenges. For example, Muhammad Adil \emph{et al.}\cite{Adil20235G6GEnabledMT} point out that industrial automation supported by 5G/6G can operate on an independent platform that supports multiple interconnected applications performing different tasks. This deployment creates a metaspace at the network edge that faces potential threats such as DoS attacks, malware attacks, and authentication attacks. (Metaspace represents a logical space that manages physical space through control-plane separation, similar to software-defined networking and other enabling technologies\cite{Khan2024AJC}). Among these threats, malware can infiltrate the metaspace through unauthorized access and execute malicious activities. To mitigate these issues, it is best to use a robust yet lightweight authentication protocol that ensures the security of both customers and manufacturers. However, as previously mentioned, while various authentication schemes have been used to address the security of edge and cloud devices in the Industrial Metaverse, they are computationally complex and communication-intensive. Designing a lightweight authentication and data preservation scheme in the context of dynamic and hierarchical models may establish secure boundaries for subnets economically and efficiently. Additionally, Trusted Execution Environment (TEE) can serve as another promising security technology to address data confidentiality, availability, and integrity issues\cite{Wang2022RTTEERS}.

Using VR was once considered relatively secure because VR devices have a closed structure that prevents external attackers from directly observing displayed content. Unfortunately, recent researchers have emphasized that virtual reality users also face serious privacy threats when confronted with side-channel attacks\cite{Zhong2023MetaverseCE}. On one hand, internal side-channel attackers can utilize motion sensor measurements to perform identity inference. For example, in XR systems, head movements play a crucial role in user interactions and navigation within virtual environments. Attackers can use machine learning techniques to analyze head movement data, associating specific movements with certain tasks or interactions, thereby reconstructing the algorithms being used and revealing sensitive information about product design and functionality. 
On the other hand, external attackers can exploit fluctuations in wireless signals emitted by VR devices to monitor password inputs\cite{Li2024DangersBC}. For instance, various systems in factories rely on identity authentication, including fingerprint recognition, facial recognition, and more. Hackers can potentially learn and reconstruct facial information to gain access to devices. Currently, the most direct countermeasure is to design new authentication mechanisms. For example, Wang \emph{et al.}\cite{Wang2022AcoPalmAP} design a palmprint-based authentication scheme that collects backscattered acoustic signals for verification, thereby mitigating privacy risks associated with image recognition. In the future, adopting hybrid multidimensional authentication technologies, supplemented by timely firmware updates, may effectively counteract privacy breaches during interactions.

Lastly, beyond the aforementioned issues, addressing security challenges in the Industrial Metaverse necessitates keeping abreast of relevant legal documents and enterprise standards.

\subsubsection{Summary}
In the above discussion, we have explored data and network security issues in the Industrial Metaverse. The security of industrial data sources at the sensor level requires mechanisms such as data masking and whitelist/blacklist strategies to reduce unnecessary data collection and mitigate privacy risks. To address malicious injection of low-quality data, solutions involving sensor node trustworthiness tracing and anomaly detection were proposed. Regarding data and network security threats in distributed industrial platforms, the decentralized architecture introduces risks of data breaches and network attacks\cite{Lin2024DecentralizedPI}. BC technology and cloud-edge collaborative strategies can enhance data management and storage security, though challenges such as high energy consumption, information overload, and the ``black box" effect remain. Finally, enabling technologies such as AI and XR also present potential security risks, including AI algorithm transparency issues and malicious exploitation, BC key management risks, and identity authentication issues in XR interactions. Addressing these challenges requires multi-dimensional technical solutions as well as supportive laws and standards.

\subsection{Resource Allocation}\label{3B}
Due to the increasing number of applications and industrial devices involved in industrial production, distributed deployment of the Metaverse platform may lead to performance imbalances among subsystems and processors, such as computing power and throughput. When facing various types of tasks throughout the PLC, enterprise demands are often dynamic and diverse, posing challenges to the management of network and computing resources. Below, we discuss resource allocation strategies in the Industrial Metaverse focusing on computing resources and storage resources.
\subsubsection{Computing Resource Allocation Issues}
The Industrial Metaverse demands significant computational resources to seamlessly render 3D virtual environments. However, due to the limited computational capabilities of certain lightweight devices within factories, they may fall short for computationally intensive rendering tasks. Although cloud servers offer sufficient computing power for remote rendering, the current state of cloud infrastructure may struggle to meet the ultra-low latency demands of industrial production activities\cite{Deng2022LowlatencyFL}.

In collaborative multi-enterprise Industrial Metaverse platforms, some adaptable traditional distributed technologies are gradually being introduced. These include MEC and FL\cite{9946372}. Al-Quraan \emph{et al.}\cite{AlQuraan2021EdgeNativeIF} discuss the potential applications of FL in various wireless networks, such as visible light communication and massive multiple-input multiple-output (mMIMO), to maintain network performance stability by training models that alleviate microbase station congestion.
Nevertheless, despite offloading computation to the edge, FL may still face challenges in frequent information exchange between cloud servers and edge nodes due to the continuous generation of large volumes of new data in industrial production processes. Additionally, the increasing number of parameters and limited bandwidth resources may become bottlenecks for FL. Furthermore, due to differences in device models and performance in factories and varying requirements for model accuracy and computational resources in different scenarios, it is difficult to consider a universal FL aggregation design solution\cite{Zhu2024TrustworthyBF}. 
Furthermore, the number of nodes in Industrial Metaverse systems is often indeterminable during application deployment, necessitating dynamic adjustments in resource allocation to accommodate the addition or removal of nodes\cite{Chen2023FederatedLF}. In the face of extreme scenarios, the robustness of the deployment architecture must also be taken into consideration.

Another common architecture is based on the allocation of computing resources in MEC\cite{Zhang2023FederatedLI}. MEC, by deploying computing nodes near the terminals, is particularly suitable for high-concurrency industrial scenarios. Since edge computing nodes are close to physical devices, they can continue local data processing and decision-making even in cases of unstable or disconnected network connections\cite{DBLP:journals/tvt/ZhangZM22}.
Tan \emph{et al.}\cite{DBLP:journals/wc/TanAWW23} introduce DT into edge networks and designed power optimization based on Deep Q-Networks (DQN) for dynamic channel conditions in MEC. The proposed architecture was validated through simulations for applications in automated warehousing logistics and mobile production lines. In Industrial Metaverse platforms, VSPs are responsible for providing resources ordered by various manufacturers. However, excessive resource allocation may lead to resource over-provisioning issues. Ng \emph{et al.}\cite{DBLP:conf/icc/NgLNXNM22} propose a decision framework for VSPs in the Metaverse based on two-stage stochastic integer programming to minimize overall network costs considering the uncertainty of user demands. However, they did not consider solutions for resource demands in the case of large-scale resource requirements.
To address these challenges, Chu \emph{et al.}\cite{DBLP:conf/wcnc/ChuNHPDNS23} decompose the Metaverse platform and utilized application similarity to improve resource utilization. They developed a self-learning framework based on Markov processes to capture implementation characteristics of the Metaverse. Considering the dynamic industrial environment, Yu \emph{et al.}\cite{DBLP:journals/tii/YuYXDLQMKC22} propose an end-to-end self-organizing resource allocation framework under a 5G heterogeneous network, greatly enhancing QoS.

In scenarios involving multiple for-profit enterprises, the utility of VSPs in the Industrial Metaverse is considered a function of latency and resource payment costs induced by resource offloading. When the utility is non-negative, VSPs will be willing to invest more computational resources in enterprise collaboration. One scheduling solution involves designing incentive mechanisms to maximize device utilization. Li \emph{et al.}\cite{Li2019CrowdBCAB} introduce a distributed crowdsourcing framework called CrowdBC, where smart contracts are used for automated execution, and BC is used for user reputation management and incentive mechanisms.
Reputation reflects the past performance of enterprise employees in task resolution. After submitting the execution plan, enterprise employees can request the system to perform task evaluation and record it. As the key to synchronizing the real world with the virtual world lies in the data collected by IoT devices and sensors, Han \emph{et al.}\cite{Han2021ADR} design a dynamic resource allocation framework using a hybrid evolutionary dynamics approach to synchronize the Metaverse with IoT services and data.
However, the aforementioned works did not consider the dynamic characteristics of each computing node in the Industrial Metaverse system, such as smart handheld terminals and AGVs. When nodes are predetermined, idle computational resources may remain underutilized when MEC handles computationally intensive tasks\cite{Yang2024TrustedME}.

\subsubsection{Storage Resource Allocation Issues}
Due to the fundamental characteristics of the Metaverse, such as diverse applications, heterogeneous connected devices, ubiquitous user access, and digital identities, in addition to computational capacity, a sustainable storage platform and efficient storage strategy are urgently needed. Servers will collaborate to store information and share content with other servers. In this sense, the payload of redundant data processing and uploading can be reduced by dynamically obtaining terminal information, predicting its demand, and collaboratively storing data, thereby improving the utilization of storage resources and reducing the service latency of the entire system.
However, while storing sensor data in the cloud may be a viable solution, as mentioned earlier, cloud-based distributed storage requires continuous internet connectivity and becomes unavailable when the platform encounters emergencies due to internet connection disruptions. Therefore, Rashid \emph{et al.}\cite{DBLP:conf/hpcc/RashidZW19} proposed the idea of using private edge devices for decentralized distributed data storage in emergencies. They developed a game theory model to dynamically incentivize resource allocation and regulate the QoS of the network. Devices contributing more storage space will receive better QoS (i.e., network bandwidth). To address edge device configuration issues, the authors developed an application that seamlessly runs in Docker containers and connects to the EdgeStore API for system configuration. EdgeStore enables maximizing system availability in emergencies by establishing cooperative storage sharing plans and storing data on private edge devices.

Storage functionality typically includes content storage and policy storage. In the policy storage scheme proposed by Zhao \emph{et al.}\cite{DBLP:conf/infocom/PoularakisLTTT19}, when the server receives relevant requests, it processes tasks based on the request and then returns them to the selected policy device terminal. In Liu \emph{et al.}'s\cite{DBLP:journals/access/LiuXYWM19} storage scheme, content is pre-stored on servers. When relevant requests arrive, the server can directly return the data to the terminal. Overall, content storage has lower latency. However, the policy storage scheme can better generalize and adapt to heterogeneous applications. It can support different data requests from users well through preset policies.

Unlike the stable resource scheduling in cloud centers, edge servers are less inclined to share storage space and data due to specific requirements such as workload, latency, and privacy. In emergencies, the inconsistent availability information of storage resources on edge servers may lead to a high data loss rate. To address this issue, Vaquero \emph{et al.}\cite{DBLP:journals/fgcs/VaqueroCEBSZ19} propose a method to replicate data on all nodes. However, the above method will increase the overall cost of the system and may degrade performance. Recent research work also involves edge caching technology in Metaverse scenarios. Cai \emph{et al.}\cite{DBLP:journals/corr/abs-2205-01944} design an efficient control policy that integrates computing, caching, and communication (3C) resources and implements the policy by implementing a multi-pipeline traffic control and 3C resource scheduling mechanism. To further optimize the online delivery performance of data-intensive services, the authors also proposed a database placement strategy based on throughput maximization and two effective database replacement strategies. On the other hand, Huynh \emph{et al.}\cite{DBLP:journals/wcl/Van-HuynhKMDD22} propose an innovative DT solution that combines MEC and URLLC to support Metaverse applications. From the perspective of the 3C integration model, this solution optimizes edge caching policies, task offloading strategies, and the allocation of computing and communication resources to address the latency optimization problem in the Metaverse enabled by DT.

\subsubsection{Summary}
The second part discusses the challenges and solutions for resource allocation strategies in the Industrial Metaverse. First, in terms of computing resource allocation, distributed technologies such as cloud computing, MEC, and FL have been introduced to optimize resource distribution due to the varying performance of applications and devices in the Industrial Metaverse. While cloud computing offers powerful computing capabilities, it struggles to meet the low-latency requirements of industrial production; MEC, on the other hand, deploys computing nodes closer to terminal devices, making it suitable for high-concurrency scenarios, although dynamic adjustments to resource allocation strategies are still required to adapt to changing demands. In addition, we explored incentive mechanisms and resource utilization optimization to improve computing resource efficiency. Secondly, in the context of storage resource allocation, dynamic storage strategies and edge storage technologies have been employed to reduce data transmission latency and duplicate processing, particularly for the diverse applications and heterogeneous devices in the Metaverse. However, inconsistency in the sharing of storage resources among edge servers remains an issue. Overall, resource allocation in the Industrial Metaverse must comprehensively consider both computing and storage needs, leveraging distributed technologies, edge computing, and dynamic strategies to enhance platform flexibility and sustainability.

\subsection{Interoperability}\label{3C}

In the context of Industry 5.0, the role of humans in the Industrial Metaverse has become increasingly important. Unlike Industry 4.0, which primarily focuses on automation and machine-driven processes, Industry 5.0 emphasizes human-machine collaboration. This collaboration not only involves interaction between humans and machines at the hardware level but also extends to cognitive cooperation. Human involvement brings unique decision-making abilities, creativity, and adaptability to the process\cite{Zhou2024PersonalizedFL}. Moreover, human-AI collaboration allows humans to focus on higher-level tasks, while AI handles repetitive or data-intensive tasks, thereby improving the efficiency and harmony of workflows. At the same time, the Metaverse platform enables human participation by fostering cultural integration and facilitating trustworthy, efficient communication, with interoperability serving as the medium to achieve all of this.
As a complex system, the Industrial Metaverse platform integrates various sensors and control subsystems with different hardware architectures and control components. If there is no support for interoperability, user identities and industrial equipment data will be limited to specific VSPs, contradicting the original intention of an interconnected Metaverse\cite{Rospigliosi2022MetaverseOS}. Therefore, interoperability of subsystems within the platform and interoperability across platforms are important considerations. 
Currently, addressing the interoperability issues in the Industrial Metaverse primarily focuses on three aspects: identity and credential authentication, HCI technology, and interface and integration frameworks.

Identity and credentials define the unique encoding of users and production equipment, while binding identifiers to attributes such as device status and personnel behavior, serving as a foundational element for achieving interoperability. This forms the foundation for achieving interoperability. These attributes can be provided by physical devices in factories or created by enterprise personnel themselves. In different platforms, the same identity and identification correspond to the same static attributes. Gadekallu \emph{et al.}\cite{DBLP:journals/fgcs/HuynhTheGWYRPCL23} suggest that BC can facilitate the exchange of data between different sub-Metaverses. They proposed using cross-chain protocols to exchange data on multiple BCs located in different virtual worlds. Therefore, Li \emph{et al.}\cite{Li2023MetaOperaAC} propose a MetaOpera, a BC-based interoperability protocol for the Metaverse. It supports cross-chain technology and on-chain/off-chain techniques for the interaction between MetaOpera and decentralized platforms as well as top centralized platforms. Compared to authentication technologies based on Sidechains\cite{DBLP:conf/sp/2019}, this method achieves lower interaction latency. Additionally, Ghirmai \emph{et al.}\cite{10189537} propose using Self-Sovereign Identity (SSI) technology, which replaces traditional identity verification with distributed identity verification using zero-knowledge proofs. For example, two workers docked on a Metaverse platform can prove each other's public keys according to the principles of SSI and encrypt their interaction records using pre-shared keys. Each worker holds a unique NFT as a unique identifier. However, the authors also pointed out that the issue of BC storage should not be ignored because controlling and managing data and identities may require different storage mechanisms\cite{Ren2025BlockchainSO}. Additionally, their scenario only applies to static attributes in the Metaverse and does not involve behavioral objects.

HCI technology, based on identity-authenticated operations, provides users with the ability to interact with virtual environments. Multimodal interaction combines various interaction methods, including motion tracking, haptic feedback, vision, and voice, offering Industrial Metaverse users a natural and intuitive interaction experience for executing complex tasks and operations\cite{FernndezCarams2024ForgingTI}. Currently, most motion tracking technologies, such as Vicon\cite{Merker2023MeasurementAO} or Xsens\cite{Debertin2024ReliabilityOX}, rely on optical markers or inertial measurement units (IMUs). However, these systems are costly to set up and require lengthy calibration times, which smaller companies may struggle to support. To address this, Ponton \emph{et al.}\cite{Pontn2023SparsePoserRF} propose a data-driven method called SparsePoser for body posture reconstruction with sparse data. This method uses six six-degree-of-freedom tracking devices to create avatars. SparsePoser encodes information obtained from sensors along with the user’s static representation and decodes it into full-body postures by reconstructing all joints between end-effectors. Jang \emph{et al.}\cite{Jang2023MOVINRM}, on the other hand, focus on reducing equipment costs and proposed a new feature encoder that learns the correlation between historical 3D point cloud data and global and local posture features, achieving real-time motion capture with a single LiDAR.
Modeling muscle activity by simultaneously detecting muscle deformation and bioelectrical signals typically requires multiple sensors or even large detection devices, which significantly reduces the wearer's comfort. To address this, Wang \emph{et al.}\cite{Wang2024ADS} introduce a fully integrated dual-mode wearable system, laying the foundation for a lightweight, dual-modal muscle sensing frontend. Suo \emph{et al.}\cite{Suo2024AIEnabledSS} present a non-invasive muscle-sensing wearable device that, when combined with machine learning, enables motion-control-based interaction between the physical body and Metaverse avatars.

Another important topic in HCI is haptic technology, as the initial goal of HCI is to enable users to perceive and interact with virtual environments in the same way they would with physical environments. Mel Slater \emph{et al.}\cite{Slater2009PlaceIA} identify two factors that influence this sense of realism: place illusion and plausibility illusion. The former is limited by the capabilities of the interaction devices themselves, while the latter depends on the content of the depicted scene. Due to the limitations in bandwidth and dynamic range, achieving physical stimulation that exactly matches human perception is often impractical. Therefore, Lee \emph{et al.}\cite{Lee2016MotionES} suggest using optical or acoustic signals to algorithmically generate vibrations. They categorized the motion into slow and fast, as well as discrete and continuous. Slow motion is relatively easy to reconstruct, but other content often requires a deeper understanding of scene semantics. Kim \emph{et al.}\cite{Kim2014SaliencyDrivenRV} provide experimental evidence to support this approach. Similarly, research on extracting vibrations from acoustic signals still requires experts for measurement or improvements in the output of suboptimal reconstruction algorithms.

In addition to motion tracking and haptic technology, HCI technology in the Industrial Metaverse also involves digital human simulation. In industrial settings, engineers often use human simulation systems, such as Jack, to assess posture, reachability, and other factors\cite{Blanchonette2010JackHM}. However, these systems typically focus on body structure and range of motion, overlooking the user’s psychophysiological state. To address this, Eyam \emph{et al.}\cite{10552271} introduce the concept of meta-states—digital representations of psychophysiological states—in industrial simulations to reflect the psychophysiological state of digital humans. This allows subjects immersed in virtual reality and professionals operating desktop-based applications to understand the state of human workers. However, due to the high individual variability in psychophysiological states, achieving this simulation poses challenges in terms of algorithmic complexity and computational resource consumption.

\begin{table*} [!t]
\begin{center}   
\caption{Research on Existing issues in the Industrial Metaverse.} 
\label{table5}
\renewcommand\arraystretch{1.3}
\begin{tabular}{|m{1.5cm}<{\centering}|m{2.5cm}<{\centering}|c|p{3.8cm}|p{6.6cm}|c|}   
\hline     
\textbf{Challenges} & \textbf{Issues} & \textbf{Ref.} & \textbf{Algorithm / Framework} &\textbf{Contribution / Performance}& \textbf{Year}\\  
\hline     
\multirow{8}{*}{\shortstack{Security \\ \& \\Privacy}} 
&\multirow{3}{*}{\shortstack{Sensor-side \\Industrial\\ Data Source}} 
&\cite{Wu2021DAPterPU}
&A data misuse prevention mechanism called DAPter. 
&DAPter can substantially raise the bar of the data abuse difficulty with little impact on QoS and overhead. 
&2021\\

\cline{3-6} 
&&\cite{Pang2022FalseDI}
&Kalman anomaly detector based on residuals.
&This may be helpful for guaranteeing the secure control of a networked system by protecting partial critical sensor measurements from FDI attacks. 
&2022\\

\cline{3-6} 
&&\cite{Guo2022DetectionOS}
& A random coding scheme to detect residuals.
&Estimation-optimization iteration algorithm is to obtain a numerical solution of the coding signal covariance.
&2022\\

\cline{2-6} 
&\multirow{2}{*}{\shortstack{Data \& Network \\Security }} 
&\cite{Peng2023DistributedIF}
& A two-stage incentive mechanism.
& Efficient resource allocation and computation offloading are realized while privacy is guaranteed.
&2023\\

\cline{3-6} 
&&\cite{Shen2021SecureSO}
& BC and DT-based distributed storage and cloud storage.
& The algorithm is better than traditional method for maximizing the total social benefits.
&2021\\

\cline{3-6} 
&&\cite{Kim2020APADAP}
& Two payload-based anomaly detection methods.
& The accuracy and the FP rate are superior to the remaining methods.
&2020\\

\cline{3-6} 
&&\cite{Kwon2020IEEE1P}
& An intrusion detection system for an IEEE 1815.1-based power system network.
& Five types of CMB attacks and three types of FDI and DR attacks are successfully detected.
&2020\\

\cline{2-6} 
&\multirow{1}{*}{\shortstack{Inherent Security \\Vulnerabilities}} 
&\cite{Wang2022AcoPalmAP}
& AcoPalm, a palm print recognition system.
& AcoPalm is resistant to replay and mimicry attacks, and can achieve 96.22\% authentication accuracy. 
&2022\\

\hline
\multirow{11}{*}{\shortstack{Resource \\ Allocation}} 
&\multirow{6}{*}{\shortstack{Computing Resource\\ Allocation Issues}} 
&\cite{DBLP:journals/wc/TanAWW23}
& A framework for MEC based on networked digital dual bases.
& Web-based DT frameworks with AI capabilities can further reduce task processing latency and improve QoS.
& 2023\\

\cline{3-6} 
&&\cite{DBLP:conf/icc/NgLNXNM22}
& A stochastic optimal resource allocation scheme based on stochastic integer programming.
& The proposed scheme minimizes the cost of VSPs while taking into account the uncertainty of user demand.
& 2022\\

\cline{3-6} 
&&\cite{DBLP:conf/wcnc/ChuNHPDNS23}
& A framework based on semi-Markov decision processes.
& The proposed approach can realize up to 120\% revenue for Metaverse VSPs and a 178.9\% probability of acceptance for Metaverse application requests.
& 2023\\

\cline{3-6} 
&&\cite{DBLP:journals/tii/YuYXDLQMKC22}
& Resource allocation mechanism for IIoT in 5G heterogeneous networks.
& It is possible to achieve better performance than other traditional deep learning (DL) methods and maintain quality of service above accepted levels.
& 2022\\

\cline{3-6} 
&&\cite{Li2019CrowdBCAB}
& A BC-based decentralized crowdsourcing framework.
& A software prototype is implemented on the Ethernet public test network with real datasets.
& 2019\\

\cline{3-6} 
&&\cite{Han2021ADR}
& A hybrid evolutionary dynamics approach.
& A dynamic resource allocation framework to synchronize the Metaverse with IoT services and data.
& 2021\\

\cline{2-6} 
&\multirow{5}{*}{\shortstack{Storage Resource\\ Allocation Issues}} 
&\cite{DBLP:conf/hpcc/RashidZW19}
& Edge-based distributed storage system.
& EdgeStore outperforms state-of-the-art distributed storage systems in terms of throughput, energy consumption, and latency.
& 2019\\

\cline{3-6} 
&&\cite{DBLP:conf/infocom/PoularakisLTTT19}
& Joint optimization of service placement and request routing with multidimensional constraints.
& The algorithm maximizes the number of requests served by the low latency edge cloud server.
& 2019\\

\cline{3-6} 
&&\cite{DBLP:journals/access/LiuXYWM19}
& Cache-enhanced multi-user MEC system.
& The proposed joint optimization of caching, computation, and communication methods can improve energy efficiency at a lower time cost.
& 2019\\

\cline{3-6} 
&&\cite{DBLP:journals/fgcs/VaqueroCEBSZ19}
& Introduced next-generation orchestration technology.
& Orchestration requirements for next-generation workflow management in microservices architectures and IoT environments are identified.
& 2018\\

\cline{3-6} 
&&\cite{DBLP:journals/corr/abs-2205-01944}
& Designed the first throughput-optimal control strategy.
& The proposed novel multi-pipeline flow control and 3C resource orchestration mechanism has excellent performance.
& 2022\\

\hline
\multicolumn{2}{|c|}{\multirow{4}{*}{\shortstack{Interoperability}} }
&\cite{Li2023MetaOperaAC}
&A protocol called MetaOpera generalized cross-Metaverse interoperability.
& The size of cross-global proofs and the average time of cross-global transactions are reduced by 8x and 3x, respectively, using the proposed solution compared to the Side-chain solution.
&2023\\ 

\cline{3-6}
\multicolumn{2}{|c|}{}
&\cite{10189537}
&Self-Sovereign Identity (SSI) integrated with BC.
& Helps address issues of decentralization, trust and interoperability in the Metaverse.
&2022\\ 

\cline{3-6}
\multicolumn{2}{|c|}{}
&\cite{Rawal2022TheRO}
&Introduceing an interoperable splitting protocol.
& The resilience and high-availability of a Metaverse system for role switching.
&2022\\ 

\cline{3-6}
\multicolumn{2}{|c|}{}
&\cite{10279406}
&Iterative algorithms for joint sub-elements and DT associations on MEC servers.
& The proposed solution coordinates the interactions between the DT and the subuniverse, thus reducing the sub-synchronization time by 25.75\%.
&2022\\ 

\hline   
\end{tabular}   
\end{center}  
\end{table*}

Additionally, during the transition from Industry 4.0 to Industry 5.0, industrial robotics technology faces the need for intelligent and highly integrated development. For industrial robots, Ma \emph{et al.}\cite{Ma2024MetaverseAM} propose application scenarios for machine-readable standards within the Metaverse, including resource retrieval, knowledge Q\&A, personalized knowledge push, and virtual digital humans. You \emph{et al.}\cite{You2024EvolutionOI} explore the implementation of virtual-real integration and human-machine collaboration within the Industrial Metaverse. They proposed a design framework for a virtual-real interaction system based on Robot Operating System (ROS) and web technologies, including the design and implementation of a general communication mechanism using ROS, enabling connectivity and real-time data communication between physical robots and virtual models through ROS topic subscriptions. Furthermore, they utilized URDF model transformation technology for model invocation and display.

Finally, as the key to system integration, interfaces and integration frameworks facilitate communication and functional collaboration between different subsystems and platforms. Split learning, as a distributed architecture, involves multiple devices and servers in an industrial platform\cite{Khan2022FederatedSL}. It can split the backbone network into multiple sub-networks, execute multiple sub-tasks in parallel on different devices or servers, and then integrate the results. Inspired by this, Rawal \emph{et al.}\cite{DBLP:conf/iri/2022} introduce a Split-Protocol that provides interoperability services. They assume that critical elements and functions are provided by a meta-server, and the system consists of ten identical meta-servers, each executing specific functions of the Metaverse. Role transitions can be achieved through the Split-Protocol. This protocol with role-switching capability can improve the availability of the Metaverse system. Furthermore, Hashash \emph{et al.}\cite{10279406} emphasize the coordination between DTs and Edge Metaverse. A Physical Twin (PT) system running in a large-scale sensing area is replicated as a Cyber Twin (CT) system on MEC servers. The area is divided into distributed subgrids and transmitted remotely to MEC servers. Their designed interactive algorithm utilizes optimal transport theory, considering both computational power and data synchronization issues.

\textit{Summary:} As a complex system, the Industrial Metaverse needs to address interoperability issues both within subsystems and across platforms. Identity and identification authentication serve as the foundation for interoperability by binding attributes such as device status and human behavior, ensuring consistency across different platforms. BC technology, cross-chain protocols, and distributed identity authentication provide technical support for data exchange and verification. Additionally, distributed architectures such as the Split Protocol, along with the coordination of DTs and edge computing, enable efficient collaboration and data synchronization within the Industrial Metaverse. However, the Industrial Metaverse platforms currently lack a unified standard development model, making it crucial to enhance interoperability and compatibility. Furthermore, only through an open collaborative ecosystem among different companies and industries can the content and value of the Industrial Metaverse be enriched, thereby improving its overall operational experience\cite{DBLP:journals/internet/JaiminiZBS22}.

\section{Standardization}\label{4}

\begin{table*}[!t]
    \begin{center}   
    \caption{Comparison of 3D Data Formats in the Metaverse}
    \label{table6}
    \renewcommand\arraystretch{1.3}
        \begin{tabular}{|m{1.5cm}<{\centering}|p{2.6cm}|p{3cm}|p{1.2cm}|p{1.6cm}|p{5cm}|} 
        \hline
        \textbf{Format} & \textbf{Developer} & \textbf{Use Cases} & \textbf{Efficiency} & \textbf{Open Source} & \textbf{Suitability for Industrial}  \\ \hline
        \textbf{VRML} & Web3D Consortium & Early VR and Web & Low  & Yes & Low \\ \hline
        \textbf{glTF 2.0} & Khronos Group & Metaverse, games, Web & High  & Yes & Medium to High (good for visualization and lightweight assets) \\ \hline
        \textbf{USD} & Pixar & Large-scale 3D content creation & Medium  & Yes & High (scalable for complex simulations and large datasets) \\ \hline
        \textbf{FBX} & Autodesk & 3D animation, game & Low & No & Medium (useful for detailed models but large file size can be a drawback) \\ \hline
        \textbf{OBJ} & Wavefront Technologies & 3D model exchange & Low & Yes & Low (limited support for complex models or interactions) \\ \hline
        \textbf{COLLADA} & Khronos Group & Cross-tool scene exchange & High & Yes & Medium (suitable for cross-platform exchanges) \\ \hline
    \end{tabular}
    \end{center}  
\end{table*}

Since the concept of the Metaverse was proposed, relevant standardization work has been continuously developed to promote its industrial deployment, ensure interoperability and compatibility between different Metaverse platforms, and reduce the cost of industrial system development and maintenance. While most existing Metaverse standards focus on commercial applications, general characteristics such as interoperability and compatibility can be adapted and expanded to meet the needs of the Industrial Metaverse, providing valuable references and guidance for its development.

From a data perspective, the focus of standardization lies primarily on platform data format standardization and the standardization of data formats between physical and virtual worlds\cite{DBLP:journals/corr/abs-2403-05205}. For instance, interoperability between industrial IoT, industrial cloud platforms, and other systems often requires a common data protocol. Standardized data formats facilitate accurate data transmission, reduce technical barriers for cross-system interactions, and enable developers to create cross-platform tools more efficiently.
Since 2008, the MPEG-V standard has defined data exchanges at the interface between the real and virtual worlds to standardize 4D sensory and virtual environments\cite{DBLP:conf/icact/2023}. This standard, comprising seven parts, starts with 23005.1, which outlines an architecture and specifies information representations for interoperability. Parts 23005.2-3 cover syntax and semantics related to sensory device and sensor functions, while parts 23005.5-6 focus on data formats and data types—essential for multimodal data scenarios in the industrial field\cite{Choi2021ACS}. 
In 2017, the Metaverse Standards Forum (MSF) recognized glTF (GL Transmission Format) 2.0 as a benchmark for 3D asset interoperability\cite{DBLP:conf/vrml/SturmSTL16}, facilitating the creation and storage of high-fidelity 3D models of industrial equipment or products, enabling efficient cross-platform and multi-context transmission and display. Subsequent data format standards like USD (Universal Scene Description) and COLLADA (Collaborative Design Activity) have further advanced cross-software data interaction and file transfer. 
Table \ref{table6} provides a concise comparison of the characteristics of these Metaverse data formats and their applicability in industrial scenarios.
In 2019, the IEEE Standards Association launched the IEEE 2888 project, which enhances MPEG-V by standardizing data formats and APIs to synchronize the virtual and physical worlds, improving sensor data collection and actuator control.
IEEE 2888 comprises five parts, including specifications for sensor interfaces, actuator interface standards, digital synchronization orchestration between networks and the physical world, VR disaster response system architecture, and evaluation methods. 
Sections 2888.1-3 provide general technologies for DTs or the Metaverse, while section 2888.4 offers specific applications, such as industrial examples like chemical incident response and fire-fighting training\cite{Yoon2021InterfacingCA}. Its layered system design integrates VR content, motion tracking, sensory feedback, and real-time data processing to deliver comprehensive and immersive emergency response training environments.

\begin{figure*}[!t]
\center{\includegraphics[width=0.95\textwidth]{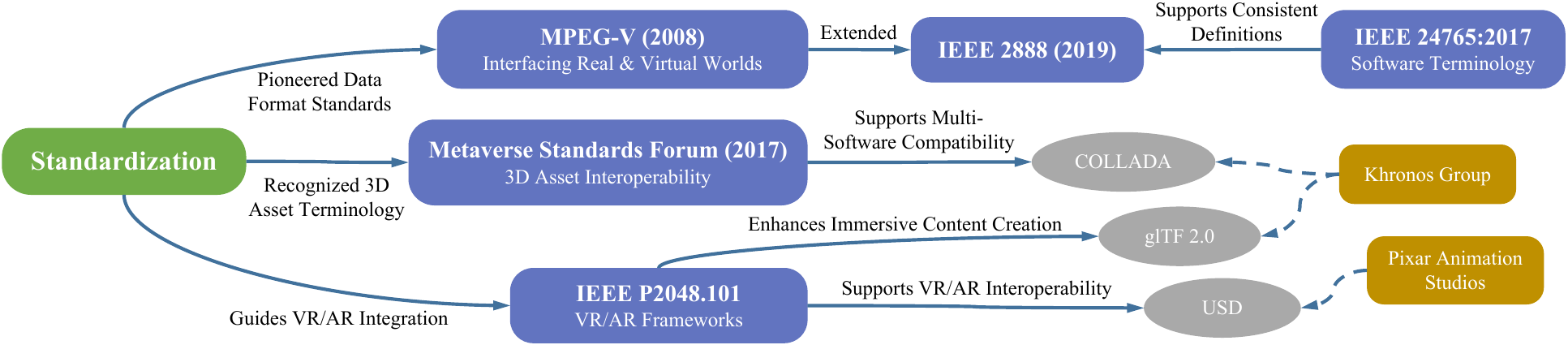}}
\caption{The logical support relationships among key standards in the standardization of the Industrial Metaverse. MPEG-V (2008) laid the foundation for interfacing real and virtual worlds and was later extended by IEEE 2888 (2019) to enhance multimodal synchronization capabilities. This was further supported by IEEE 24765:2017, which ensures consistent terminology across systems. The Metaverse Standards Forum (2017) emphasizes 3D asset interoperability and supports the following data formats: COLLADA (developed by Khronos Group, enabling multi-software compatibility), glTF 2.0 (also provided by Khronos, facilitating efficient and immersive content transmission), and USD (created by Pixar Animation Studios, enabling high-fidelity scene descriptions and cross-platform interactions). Additionally, IEEE P2048.101 offers a framework for VR/AR systems, guiding the integration of immersive technologies. 
}
\label{fig9}
\end{figure*}

From a technical perspective, ISO/IEC/IEEE 24765:2017 defines Metaverse interoperability as the ability of systems and components to exchange and use information\cite{8016712}. This includes seamless collaboration between agents, system-level data transfer, and object communication and exchange of actionable information.
In June 2022, the Khronos Group initiated and established the MSF, bringing together industry giants such as Meta, Microsoft, NVIDIA, and others across sectors like chip manufacturing and gaming. The Forum has amassed over 2,400 companies and standardization organizations, aiming to foster open Metaverse development, interoperability testing, and the deployment of open-source tools. In July 2022, the ISO/IEC JTC 1/SC 29 MPEG WG2 began collecting Metaverse case studies and conducting technical analyses to identify coding and system standards required to support future Metaverse experiences. In January 2023, the IEC Standardization Evaluation Group 15 on Metaverse (SEG 15) was established to formulate a Metaverse standardization roadmap, establish connections with related organizations, and propose further work recommendations. Recently, the IEEE P2048.101 series standards have been developed to provide technical frameworks, components, and main business processes for AR systems on mobile devices in industrial settings\cite{10154541}.
Fig \ref{fig9} illustrates the logical support relationships among key standards in the standardization of the Industrial Metaverse.

Overall, the development of Industrial Metaverse standards is still at an early stage. One early example was IBM and Second Life’s effort in 2008, which established a framework for virtual world interoperability using the Open Grid protocol\cite{Feldberg2008VUS}. However, the pressing issue for interoperability extends beyond connecting virtual worlds to encompass the evolution of the entire network, as it transitions to three-dimensional interfaces. While Abilkaiyrkyzy \emph{et al.}\cite{DBLP:journals/access/AbilkaiyrkyzyELE23} have conducted preliminary studies on small components within industrial platforms, comprehensive and systematic research remains a pressing need.

Given that industrial scenarios involve integrating various sensors and industrial control devices, existing interoperability and compatibility standards need further expansion and refinement, with higher flexibility and modular design introduced to meet diverse industrial scenarios and industry needs. As the Industrial Metaverse gradually materializes, its accompanying standards in interoperability, security and privacy, openness, scalability, and cross-industry collaboration will become increasingly robust, forming a comprehensive standard system.

\section{Future trends and outlook for Industrial Metaverse}\label{5}

While researchers have made significant efforts in studying the Industrial Metaverse tailored to industrial scenarios, alongside notable achievements, there are additional issues warranting consideration. In this section, we further expand our perspective and propose several future research directions for the benefit of researchers. Additionally, for each direction, we provide a brief list of skills/mathematical tools researchers should master, serving as foundational requirements for conducting related research:
\subsection{Flexible Communication Network Architecture}

The seamless internet connectivity of the Metaverse largely relies on reliable broadband services, with holographic communication requiring real-time operation, sufficient bandwidth, and throughput\cite{Braud2022ScalingupAU}. While low latency and high reliability are the primary advantages of 5G/6G services, consistent latency requires a flexible end-to-end architecture. This is because the Industrial Metaverse requires real-time communication of video, audio, and data. Currently, there is limited research on the network architecture and signal processing aspects of 6G in the context of the Metaverse.
This calls for a strong foundation in wireless communication theory, signal processing, and expertise in network architecture design.

\subsection{Privacy and Security Issues}

In Section \ref{3}, we have extensively discussed the confidentiality and security issues of the Industrial Metaverse from the perspectives of data, network, distributed platforms, and enabling technologies, as well as existing solutions. It can be foreseen that, as the Metaverse platform gradually matures, its security measures will largely rely on continuous upgrades through security patches. However, with the increasing prevalence of sophisticated cyber attacks, such security measures may bring challenges in terms of operational costs and scalability\cite{HuynhThe2022ArtificialIF}. In the future, it is possible to adopt intrinsic security solutions based on knowledge in the fields of quantum key distribution, cryptography, and network security, such as quantum key distribution, which utilizes channel-based keys generated through quantum entanglement to address information leakage in wireless transmissions.

\subsection{Lightweight and Ease of Use}

Since Industrial Metaverse scenarios are constantly producing, storing, and transmitting data, they require massive access to hardware devices. These devices include production equipment deployed in factories, industrial sensor devices, and access devices such as VR glasses for user interaction. The lightweighting at the hardware level has been slow to materialize. Ways to reduce hardware overhead as well as increase the flexibility and scalability of devices is a valuable topic\cite{Aburbeian2022ATA}.
Therefore, how to achieve effective support for these requirements through refined embedded system design, optimized hardware performance, and improved sensor technology is the direction that researchers need to focus on.

\subsection{Model Accuracy and Completeness}

The key factor affecting model accuracy in assembly line mode manufacturing applications is the quality of the data collected. The user must provide the system with complete and accurate available data\cite{Park2022IdentifyingWT}. Although the addition of DT technology can realize virtual samples directly into the model, the data sources often have serious heterogeneity due to different equipment models and data types, and the data fusion and calibration methods at this stage still need to be improved.
To better improve the performance of the models, researchers need to acquire skills related to industrial data preprocessing, data fusion and machine learning algorithms.

\subsection{Unified Architecture}

Central to the realization of the Metaverse lies the crucial foundation of Web 3.0 architecture. However, the intricate task of amalgamating diverse technologies within a unified architecture necessitates comprehensive efforts in web protocol design, distributed systems architecture, and web technologies. In order to address this challenge, the development of common protocols capable of accommodating the distinct needs and requirements of various industrial sectors becomes imperative. Such protocols would serve as the bedrock for seamless integration, enabling interoperability and synergy among different technologies. By fostering a standardized framework, the development of the Metaverse can progress cohesively, accommodating a multitude of industries and catering to the unique demands of users across diverse domains\cite{Suh2022UtilizingTM}.

\subsection{Decentralization in Industrial Systems}

Although systems in the Industrial Metaverse are typically not fully decentralized—some distributed components are often owned and managed by central authorities—in certain industrial scenarios, such as edge computing and multi-party collaboration platforms, decentralization can be used to reduce latency or enhance system resilience. The core feature of Web3.0 is decentralization, but current APIs are not well integrated with Web3.0. Therefore, in specific scenarios (such as supply chain management, distributed energy management systems, etc.), new decentralized protocols need to be developed to achieve the decentralization of Metaverse components\cite{GolfPape2022EmbracingFT}. While some components can be deployed in a distributed manner, the existing infrastructure relies on traditional controls and protocols, which means that to achieve decentralization, significant changes to the backbone architecture are necessary\cite{Han2022VirtualRC}. Therefore, the focus of future research is on designing decentralized protocols suitable for the Web and integrating open standards and APIs with Web 3.0 engines to enable seamless data transfer to different endpoints\cite{10468659}.

\subsection{Network Isolation Issue}

Although OT (Operational Technology) and IT (Information Technology) networks have been integrated, they still exist in isolated silos. OT and IT systems typically use different technology stacks and protocols. OT systems are mostly specialized industrial control systems, while IT systems focus more on information storage, processing, and exchange. These technological differences create significant challenges in interoperability, data transmission, and management when integrating OT and IT networks. Although Szucs \textit{et al.}\cite{Szucs2023DataIF} provide a use case that collects data from OT/IT systems and prevents interface confusion, thus simplifying the enterprise network architecture, the issues of coordinating between different technology stacks, protocols, and security requirements, as well as the specific implementation details, remain to be addressed.

\subsection{Efficient Cross-chain Authentication}

Efficient cross-chain authentication and governance is crucial to ensure the security and legitimacy of digital asset-related activities (e.g., asset transactions) across different submetadomains built on heterogeneous BCs. Current cross-chain mechanisms are mainly focused on digital asset transfers and rely on notary schemes, hash locks, relay chains and side chains, with few mechanisms considering cross-chain authentication and governance for meta-domains. The implementation, efficiency and security of cross-domain and BC authentication need further research. Moreover, novel decentralized, hierarchical and penetrating cross-chain governance mechanisms also need further research in the Metaverse. In addition, efficient meta-space-specific consensus mechanisms, redesigned block structures, and well-designed user incentives are all necessary for unique meta-space applications\cite{Zhang2022TheMI}. Achieving these goals requires a deep understanding of BC cross-chain technologies, authentication protocols, and the ability to program smart contracts in order to design and implement scalable cross-chain governance architectures, dynamically collaborative governance rules, and on-chain entity identification and risk assessment mechanisms.

\subsection{Integration with Large-scale Pre-trained Models}

Large-scale pre-trained models hold the potential to offer exquisitely tailored and individualized encounters within the Metaverse. Through discerning analysis of user inclinations, behaviors, and historical data, these models can be refined to generate bespoke virtual panoramas, characters, and interactions, ushering in a realm of heightened immersion and personalized engagement within the Metaverse.

Nevertheless, notable challenges persist. One such challenge lies within the realm of the language macrocosm, where Metaverse users hail from diverse nations and regions, conversing in various languages. Consequently, it becomes imperative for the model to possess the capability to seamlessly interact across linguistic boundaries. 
Additionally, ensuring the seamless deployment, high concurrency, and stability of the model is essential to quickly respond to and handle user requests\cite{Ding2023ParameterefficientFO}. Addressing these challenges requires expertise in natural language processing, deep learning model training, and skills in model deployment and optimization to support multilingual interactions and efficient user engagement within the Metaverse.

\subsection{Integration with Quantum Computing}

Quantum computing, with its unparalleled capacity to handle intricate calculations and tackle problems exponentially faster, possesses the potential to reshape the capabilities of the Metaverse and elevate its performance to unprecedented heights. A recent study conducted by Cui \emph{et al.}\cite{Cui2022ACP} delves into the implementation of secure communication and efficient cross-chain protocols utilizing quantum computing. Meanwhile, Ren \emph{et al.}\cite{Ren2022QuantumCL} have explored the utilization of game-based quantum collective learning and many-to-many matching schemes within the Metaverse, enabling the attainment of optimal strategies for maximizing system revenue.Looking ahead, quantum computing is poised to find widespread application in enhancing data processing, encryption, and security performance within the Metaverse. Furthermore, it holds the promise of expediting virtual and AR experiences, thereby propelling the Metaverse into a realm of heightened immersion and seamless interactivity.
Advancing these innovations requires a strong grasp of quantum computing theory, quantum algorithm design, and quantum communication protocols.

\subsection{Availability of Datasets for Industrial Metaverse}

In industrial environments, there is a large amount of data of different types, formats, and sources, which are often distributed across various systems and devices, leading to data isolation and fragmentation. The lack of unified data integration and management mechanisms hinders the seamless consolidation and utilization of data. Moreover, different stakeholders in industrial settings may have varying data ownership and access rights. Therefore, establishing appropriate data ownership and access control mechanisms based on theories of data management, data integration, and access control is crucial to ensuring data legitimacy and compliance.

\subsection{Ethical and Social Impacts}
When machine decisions impact people's lives, particularly in industrial environments with high demands on cost and safety, it is crucial to define the level of autonomy granted to algorithms within the system. Beyond technical considerations, the ethical use of data and the fair distribution of rights in the Industrial Metaverse must also be addressed, especially in terms of corporate social responsibility \cite{DBLP:journals/aiethics/BenjaminsVA23}. Large corporations are more likely to concentrate digital resources for advanced digital twin modeling, while the high cost of hardware may exclude smaller enterprises with limited budgets, thus exacerbating the digital divide.

With the advancement of automation technologies, AI-driven robots and virtual simulation systems will significantly reduce the need for human operators and traditional on-site workers. While this enhances production efficiency, it could also lead to the gradual obsolescence of physical roles such as machine operators, technicians, and on-site inspectors. Moreover, as industrial processes shift to digital environments, workers in traditional industries may face challenges in adapting to new skill demands, such as operating VR-based simulations, AI analytics, and digital twin technologies.

In the B2C context, protecting user privacy, data security, and combating misinformation are of critical importance. Companies should also promote diversity and inclusion, focus on environmental sustainability, and strike a balance between innovation and social responsibility\cite{UwhejevweTogbolo2024EthicalUO}. Further AI regulations should be introduced to ensure that the Industrial Metaverse operates within legal frameworks, safeguarding users' rights.

\section{Conclusion}\label{6}

In this paper, we conducted comprehensive research on the Industrial Metaverse. By analyzing the characteristics of Metaverse technology, we put forward convincing reasons for deploying Metaverse platforms in the industrial field. Furthermore, we summarized the research progress of several key enabling technologies in the Industrial Metaverse and elucidated the advantages of each technology in various stages of industrial production. We provided a comprehensive summary of various technologies, including existing challenges and performance requirements in industrial scenarios.
Furthermore, starting from the Metaverse itself and combining with the characteristics of industrial scenarios, we identified three key challenges: security and confidentiality issues, resource allocation problems under resource constraints, and interoperability. In addition, we compiled the standardization efforts in the field of Metaverse over the past few years. Lastly, we concluded with a forward-looking discussion, considering the current challenges and future research directions. 
With the continuous advancement of Metaverse-related technologies and their further integration with industrial scenarios, we anticipate that our work can provide references for future research endeavors.

\bibliographystyle{IEEEtran}
\bibliography{IEEEabrv,Ref}

\begin{IEEEbiography}[{\includegraphics[width=1in,height=1.25in,clip,keepaspectratio]{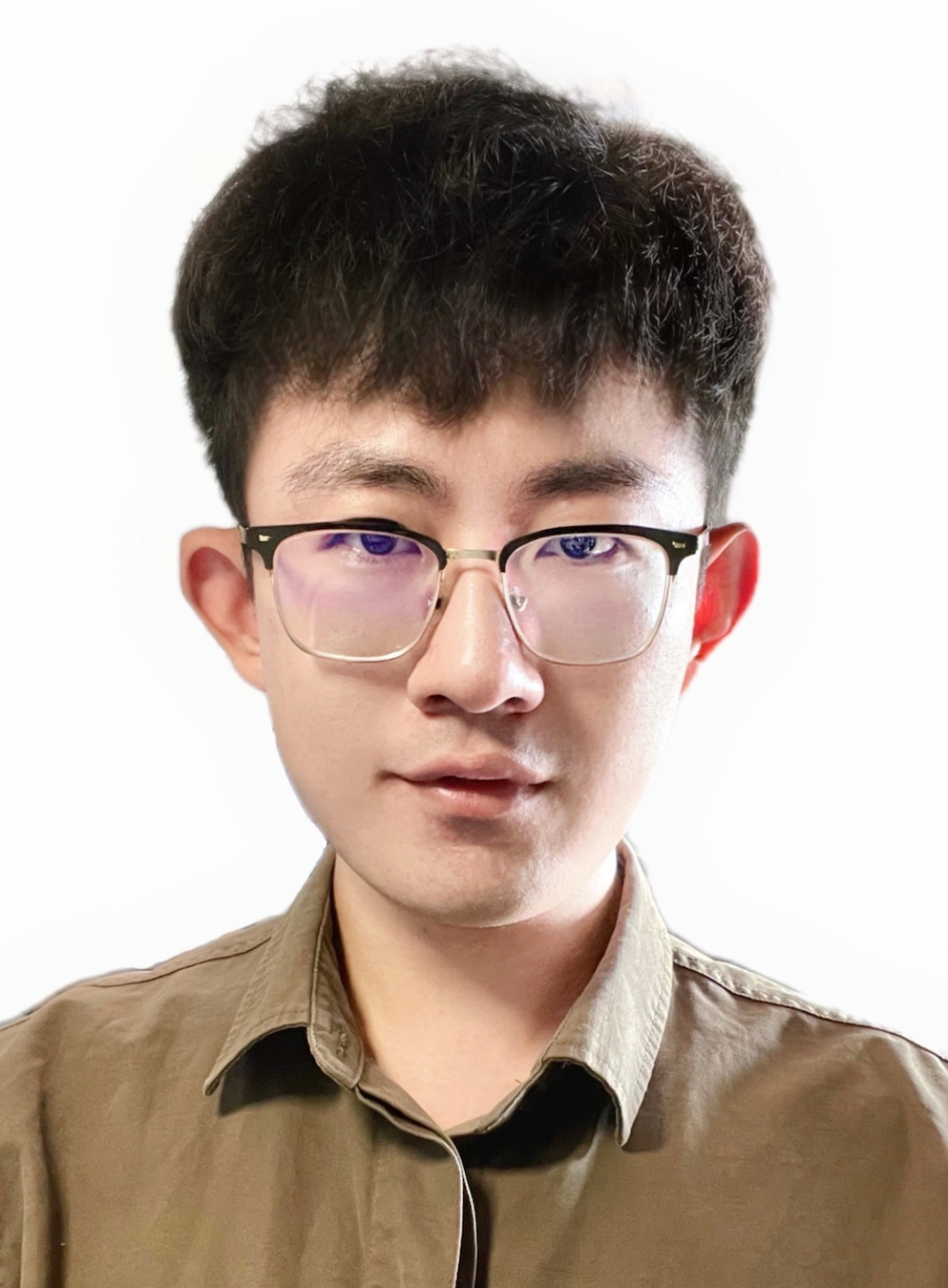}}]{Shiying Zhang}
is currently pursuing the Ph.D. degree in communication and information system with the School of Electrical and Optical Engineering, Nanjing 210094, China. His research interests focus on industrial big data analytics, generative artificial intelligence, and federated learning.
\end{IEEEbiography}
\begin{IEEEbiography}[{\includegraphics[width=1in,height=1.25in,clip,keepaspectratio]{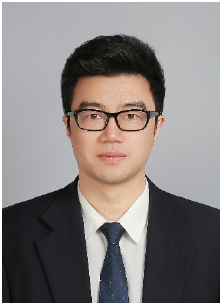}}]{Jun Li}
 (M’09-SM’16-F’25) received Ph. D degree in Electronic Engineering from Shanghai Jiao Tong University, Shanghai, P. R. China in 2009. From January 2009 to June 2009, he worked in the Department of Research and Innovation, Alcatel Lucent Shanghai Bell as a Research Scientist. From June 2009 to April 2012, he was a Postdoctoral Fellow at the School of Electrical Engineering and Telecommunications, the University of New South Wales, Australia. From April 2012 to June 2015, he was a Research Fellow at the School of Electrical Engineering, the University of Sydney, Australia. From June 2015 to June 2024, he was a Professor at the School of Electronic and Optical Engineering, Nanjing University of Science and Technology, Nanjing, China. He is now a Professor at the School of Information Science and Engineering, Southeast University, Nanjing, China. He was a visiting professor at Princeton University from 2018 to 2019. His research interests include network information theory, game theory, distributed intelligence, multiple agent reinforcement learning, and their applications in ultra-dense wireless networks, mobile edge computing, network privacy and security, and industrial Internet of Things. He has co-authored more than 300 papers in IEEE journals and conferences. He was serving as an editor of IEEE Transactions on Wireless Communication and TPC member for several flagship IEEE conferences.
\end{IEEEbiography}
\begin{IEEEbiography}[{\includegraphics[width=1in,height=1.25in,clip,keepaspectratio]{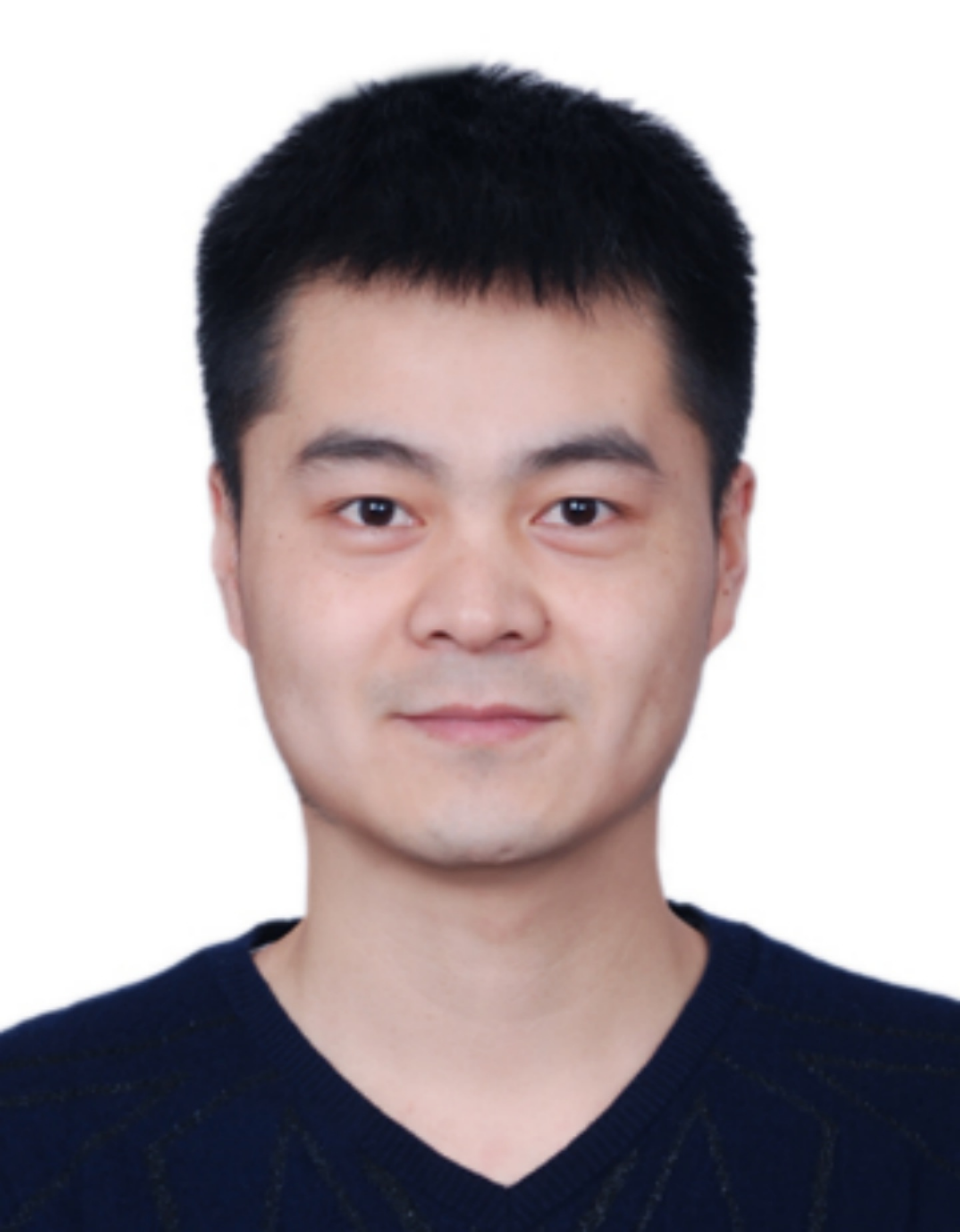}}]{Long Shi} 
(Senior Member, IEEE) received the Ph.D. degree in Electrical Engineering from the University of New South Wales, Sydney, Australia, in 2012. From 2013 to 2016, he was a Postdoctoral Fellow at the Institute of Network Coding, Chinese University of Hong Kong, China. From 2014 to 2017, he was a Lecturer at Nanjing University of Aeronautics and Astronautics, Nanjing, China. From 2017 to 2020, he was a Research Fellow at the Singapore University of Technology and Design. Now he is a Professor at the School of Electronic and Optical Engineering, Nanjing University of Science and Technology, Nanjing, China. His research interests include wireless communications, decentralized security, and edge intelligence. He is serving as an editor of IEEE Transactions on Cognitive Communications and Networking.

\end{IEEEbiography}
\begin{IEEEbiography}[{\includegraphics[width=1in,height=1.25in,clip,keepaspectratio]{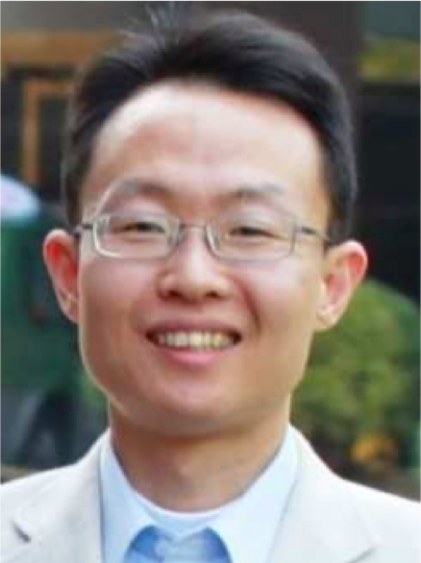}}]{Ming Ding}
(IEEE M’12-SM’17) received the B.S. (with first-class Hons.) and M.S. degrees in electronics engineering from Shanghai Jiao Tong University (SJTU), Shanghai, China, and the Doctor of Philosophy (Ph.D.) degree in signal and information processing from SJTU, in 2004, 2007, and 2011, respectively. From April 2007 to September 2014, he worked at Sharp Laboratories of China in Shanghai, China as a Researcher/Senior Researcher/Principal Researcher. Currently, he is a Principal Research Scientist at Data61, CSIRO, in Sydney, NSW, Australia. His research interests include data privacy and security, machine learning and AI, and information technology. He has co-authored more than 200 papers in IEEE journals and conferences, all in recognized venues, and around 20 3GPP standardization contributions, as well as two books, i.e., “Multi-point Cooperative Communication Systems: Theory and Applications” (Springer, 2013) and “Fundamentals of Ultra-Dense Wireless Networks” (Cambridge University Press, 2022). Also, he holds 21 US patents and has co-invented another 100+ patents on 4G/5G technologies. Currently, he is an editor of IEEE Transactions on Wireless Communications and IEEE Communications Surveys and Tutorials. Besides, he has served as a guest editor/co-chair/co-tutor/TPC member for multiple IEEE top-tier journals/conferences and received several awards for his research work and professional services, including the prestigious IEEE Signal Processing Society Best Paper Award in 2022.
\end{IEEEbiography}
\begin{IEEEbiography}[{\includegraphics[width=1in,height=1.25in,clip,keepaspectratio]{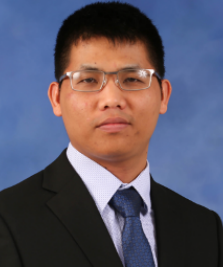}}]{Dinh C. Nguyen}
(Member, IEEE) is an assistant professor at the Department of Electrical and Computer Engineering, The University of Alabama in Huntsville, USA. He worked as a postdoctoral research associate at Purdue University, USA from 2022 to 2023. He obtained the Ph.D. degree from Deakin University, Australia in 2021. His research interests include wireless networking, distributed learning, security and privacy. He has published over 30 papers on top-tier IEEE/ACM conferences and journals. He is an Associate Editor of the IEEE Open Journal of the Communications Society.
\end{IEEEbiography}
\begin{IEEEbiography}[{\includegraphics[width=1in,height=1.25in,clip,keepaspectratio]{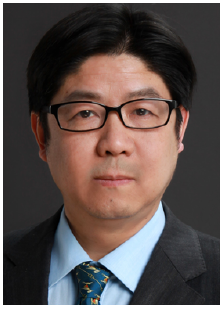}}]{Wen Chen}
(M’03–SM’11) received BS and MS from Wuhan University, China in 1990 and 1993 respectively, and PhD from University of Electro-communications, Japan in 1999. He is now a tenured Professor with the Department of Electronic Engineering, Shanghai Jiao Tong University, China, where he is the director of Broadband Access Network Laboratory. He is a fellow of Chinese Institute of Electronics and the distinguished lecturers of IEEE Communications Society and IEEE Vehicular Technology Society. He is the Shanghai Chapter Chair of IEEE Vehicular Technology Society, a vice president of Shanghai Institute of Electronics, Editors of IEEE Transactions on Wireless Communications, IEEE Transactions on Communications, IEEE Access and IEEE Open Journal of Vehicular Technology. His research interests include multiple access, wireless AI and RIS communications. He has published more than 200 papers in IEEE journals with citations more than10,000 in Google scholar.
\end{IEEEbiography}
\begin{IEEEbiography}[{\includegraphics[width=1in,height=1.25in,clip,keepaspectratio]{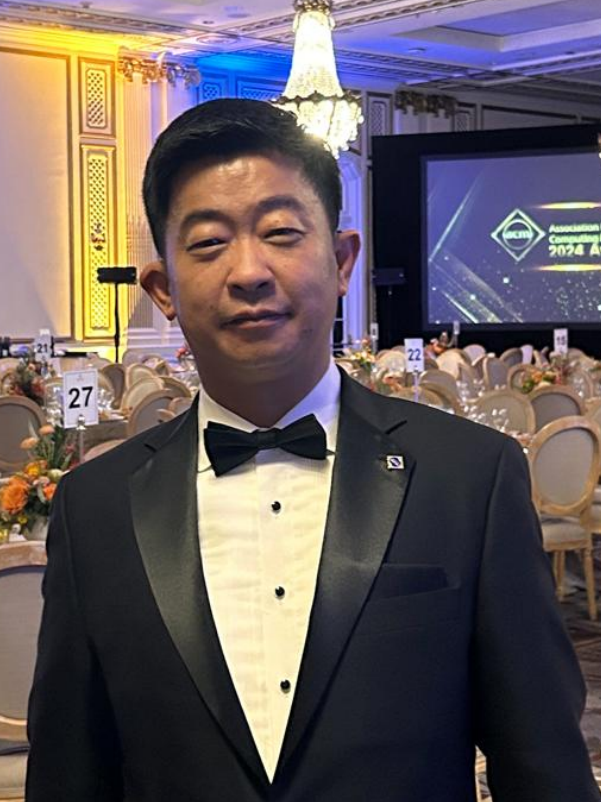}}]{Zhu Han}
(S’01–M’04-SM’09-F’14) received the B.S. degree in electronic engineering from Tsinghua University, in 1997, and the M.S. and Ph.D. degrees in electrical and computer engineering from the University of Maryland, College Park, in 1999 and 2003, respectively. 

From 2000 to 2002, he was an R\&D Engineer of JDSU, Germantown, Maryland. From 2003 to 2006, he was a Research Associate at the University of Maryland. From 2006 to 2008, he was an assistant professor at Boise State University, Idaho. Currently, he is a John and Rebecca Moores Professor in the Electrical and Computer Engineering Department as well as in the Computer Science Department at the University of Houston, Texas. Dr. Han’s main research targets on the novel game-theory related concepts critical to enabling efficient and distributive use of wireless networks with limited resources. His other research interests include wireless resource allocation and management, wireless communications and networking, quantum computing, data science, smart grid, carbon neutralization, security and privacy.  Dr. Han received an NSF Career Award in 2010, the Fred W. Ellersick Prize of the IEEE Communication Society in 2011, the EURASIP Best Paper Award for the Journal on Advances in Signal Processing in 2015, IEEE Leonard G. Abraham Prize in the field of Communications Systems (best paper award in IEEE JSAC) in 2016, IEEE Vehicular Technology Society 2022 Best Land Transportation Paper Award, and several best paper awards in IEEE conferences. Dr. Han was an IEEE Communications Society Distinguished Lecturer from 2015 to 2018 and ACM Distinguished Speaker from 2022 to 2025, AAAS fellow since 2019, and ACM Fellow since 2024. Dr. Han is a 1\% highly cited researcher since 2017 according to Web of Science. Dr. Han is also the winner of the 2021 IEEE Kiyo Tomiyasu Award (an IEEE Field Award), for outstanding early to mid-career contributions to technologies holding the promise of innovative applications, with the following citation: ``for contributions to game theory and distributed management of autonomous communication networks."

\end{IEEEbiography}

\end{document}